\documentclass[xxxx,nonblindrev]{xxxxxxxx}
\OneAndAHalfSpacedXII % current default line spacing
\usepackage{url}
\usepackage[hidelinks]{hyperref}
\usepackage[utf8]{inputenc}
\usepackage{graphicx}
\usepackage{amsmath}
\usepackage{amssymb}
\usepackage{mathrsfs}
\usepackage{booktabs}
\usepackage[switch]{lineno}
\usepackage[linesnumbered,ruled,vlined]{algorithm2e}
\usepackage{multirow}
\usepackage{float}
\usepackage[medium,compact]{titlesec}
\usepackage{setspace}
\usepackage{framed}
\usepackage{comment}

\newtheorem{reduce}{Reduction}

\newtheorem{observation}{Observation}
\newcommand{\tabincell}[2]{\begin{tabular}{@{}#1@{}}#2\end{tabular}}

\usepackage{natbib}
 \bibpunct[, ]{(}{)}{,}{a}{}{,}%
 \def\bibfont{\small}%
\TheoremsNumberedThrough     % Preferred (Theorem 1, Lemma 1, Theorem 2)
\EquationsNumberedThrough    % Default: (1), (2), ...
\MANUSCRIPTNO{XXXX} % When the article is logged in and DOI assigned to it,
\begin{document}
%%%%%%%%%%%%%%%%

%\setcounter{page}{1}
%\DOI{10.1287/xxxx.0000.0000}%

\RUNAUTHOR{Wang et al.}

\RUNTITLE{
Maximum $k$-Plex Parameterized by Degeneracy Gaps 
}

% \TITLE{Bundling Information Goods of Decreasing Value}
\TITLE{
Fast Maximum $k$-Plex Algorithms Parameterized by Small Degeneracy Gaps 
}

\ARTICLEAUTHORS{%
\AUTHOR{Zhengren Wang, Yi Zhou \footnote{Corresponding author.}, Chunyu Luo, Mingyu Xiao}

\AFF{School of Computer Science and Engineering, University of Electronic Science and Technology of China, Chengdu 611731, China, 
\EMAIL{zr-wang@outlook.com},
\EMAIL{zhou.yi@uestc.edu.cn},
\EMAIL{Chunyu-Luo@std.uestc.edu.cn},
\EMAIL{myxiao@uestc.edu.cn}
}
\AUTHOR{Jin-Kao Hao}
\AFF{LERIA, Universit$\acute{e}$ d'Angers, 2 Boulevard Lavoisier, 49045 Angers, France, 
\EMAIL{jin-kao.hao@univ-angers.fr}
}
} % end of the block
% LERIA, Universit$\acute{e}$ d'Angers, 2 Boulevard Lavoisier, 49045 Angers, France

\ABSTRACT{%
Given a graph, a $k$-plex is a set of vertices in which each vertex is not adjacent to at most $k-1$ other vertices in the set. 
The maximum $k$-plex problem, which asks for the largest $k$-plex from the given graph, is an important but computationally challenging problem in applications such as graph mining and community detection. 
So far, there are many practical algorithms, but without providing theoretical explanations on their efficiency.
We define a novel parameter of the input instance, $g_k(G)$, the gap between the degeneracy bound and the size of the maximum $k$-plex in the given graph, and present an exact algorithm parameterized by this $g_k(G)$, which has a worst-case running time polynomial in the size of the input graph and exponential in $g_k(G)$.
In real-world inputs, $g_k(G)$ is very small, usually bounded by $O(\log{(|V|)})$, indicating that the algorithm runs in polynomial time.
We further extend our discussion to an even smaller parameter $cg_k(G)$, the gap between the community-degeneracy bound and the size of the maximum $k$-plex, and show that without much modification, our algorithm can also be parameterized by $cg_k(G)$.    
To verify the empirical performance of these algorithms, we carry out extensive experiments to show that these algorithms are competitive with the state-of-the-art algorithms. 
In particular, for large $k$ values such as $15$ and $20$, our algorithms dominate the existing algorithms.
Finally, empirical analysis is performed to illustrate the effectiveness of the parameters and other key components in the implementation.
%that tests with large $k$ values such as $15$ and $20$ are overlooked in existing work
}%

%\KEYWORDS{deterministic inventory theory; infinite linear programming duality; 
%  existence of optimal policies; semi-Markov decision process; cyclic schedule}

\HISTORY{}

\maketitle
% Text of your paper here
\section{Introduction}

A \emph{clique} of a graph is a set of vertices that are pairwise connected.
The maximum clique problem (MCP), which asks for the largest clique from the given graph, is a fundamental NP-hard problem.
Applications of MCP include coding theory, computer vision, and multi-agent systems (see \cite{wu2015review,tovsic2004maximal}).
However, for many other applications such as complex network analysis, where dense, not necessarily fully connected structures are of particular interest, the clique model is over-restrictive, as shown in \cite{pattillo2012clique,pattillo_clique_2013}.
Hence, the $k$-\emph{plex} is proposed as a relaxed form of clique in \cite{seidman1978graph}.
A $k$-plex is a vertex set that is almost a clique, but each vertex of the $k$-plex is allowed to have at most $k-1$ missing adjacent vertices in this vertex set, $k$ being a positive integer. When $k$ is equal to 1, a $k$-plex is a clique.
As a basic problem of the $k$-plex model, the maximum $k$-plex problem asks for the largest $k$-plex in a given graph.
Algorithms for this problem are important tools in data analysis tasks such as community detection \citep{conte2018d2k,zhou2020enumerating,zhu2020community}, protein interaction prediction \citep{yu2006predicting} and follower-boosting on social media \citep{hooi2020telltail}.

In terms of complexity, the maximum $k$-plex problem is NP-hard \citep{yannakakis1978node}.
It is also W[1]-hard with respect to a given solution size for any fixed $k\ge 1$ \citep{khot2002parameterized,komusiewicz2009isolation}, that is, there is little hope of finding an algorithm that runs in time polynomial in the size of the input graph and exponential in its given solution size. 
Furthermore, there is no polynomial-time algorithm that approximates the optimal solution within a factor better than $O(n^{\epsilon})$ for any $\epsilon > 0$, unless P=NP \citep{lund1993approximation}.

The maximum $k$-plex problem is difficult in complexity theory, yet in recent years there have been a variety of exact algorithms that have been proven to be effective in practice. For instance,
BS (\citeauthor{xiao2017fast} \citeyear{xiao2017fast}), BnB (\citeauthor{gao2018exact} \citeyear{gao2018exact}), Maplex (\citeauthor{zhou2021improving} \citeyear{zhou2021improving}), KpLeX (\citeauthor{jiang2021new} \citeyear{jiang2021new}) and kPlexS (\citeauthor{chang2022efficient} \citeyear{chang2022efficient}). 
%Among them, Maplex, kpLex and kPlexS are the state-of-the-art algorithms. 
These algorithms often solve large sparse graphs efficiently in practice, even if the graph has millions of vertices and edges. 

It is clearly a challenging task to understand why such exact, and in worst-case, exponential-time algorithm can solve the large instance in just seconds. A number of successful attempts have been made from the perspective of parameterized complexity.  
A problem is said to be \emph{fixed-parameterized tractable} with respect to the parameter $s$ if the problem can be solved in time $O(f(s){|V|}^c)$  where $|V|^c$ represents a polynomial function of the vertex number of the input graph and $f$ is a computable function of $s$.
In expectation, if a $s$ is a small parameter, then the problem could be solved efficiently, as the complexity is nearly polynomial in this case. 
Regarding the maximum $k$-plex problem, it is parameterized by the \emph{degeneracy} of the input graph $G$, a parameter often used to measure the sparsity of $G$ \citep{komusiewicz2016multivariate}.
For large sparse graphs $G$, the degeneracy $d(G)$ is often much smaller than the number of vertices.
\citet{wang2022listing} empirically validated that if the input graph has a small degeneracy value, then the corresponding parameterized algorithm is efficient. 
However, this theoretical complexity still has limitations in explaining practical performance. It is observed that for many graphs where $d(G)$ is large, the algorithm is still very efficient.
For example, the degeneracy of the popular social network graphs soc-livejournal and scc\_reality is  213 and 1235, respectively, but existing algorithms always solve them in seconds.

In the paper, we try to fill this gap by studying a new parameterized complexity that better captures the hardness of the input instance. 
For the first time, we introduce the degeneracy gap for the maximum $k$-plex problem.
The concept of the degeneracy gap was first suggested by \cite{walteros2020maximum} for the maximum clique problem. It is expressed as $d(G)+1-\omega(G)$, with $d(G)$ being the degeneracy and $\omega(G)$ the maximum clique size of the input graph.
The degeneracy gap can also be interpreted as the gap between the upper bound of the maximum clique size $d(G)+1$ (called the degeneracy bound in the paper) and the clique size $\omega(G)$.

Extending the degeneracy gap for the maximum $k$-plex problem,  it is simply defined as $d(G)+k-\omega_k(G)$ because $d(G)+k$ is also an upper bound of the maximum $k$-plex size $\omega_k(G)$.
%Compared to degeneracy, this degeneracy gap better captures the hardness of the input instance.
For any input instance, the degeneracy gap is bounded by the degeneracy because $\omega_k(G)\ge k$.
Indeed, when the input instances are easily solvable, the degeneracy gap is often much smaller than th degeneracy.
For example, the degeneracy gap for the aforementioned graphs soc-livejournal and scc\_reality  is $1$ when $k=2$.
%\cite{walteros2020maximum} demonstrated that the maximum clique problem is parameterized by $d(G)+1-\omega_1(G)$.
Therefore, it is natural to ask the following two questions: Can the maximum $k$-plex problem be parameterized by the degeneracy gap for any fixed $k$?  If so, is there a significant correlation between parameter size and practical efficiency? Fortunately, we answer these problems affirmatively in the paper.

\subsection{Contributions}

In this paper, we continue the effort for fast maximum $k$-plex algorithms with additional guarantees that the algorithms are parameterized by smaller parameters. \footnote{A preliminary version of this paper titled “A Fast Maximum $k$-Plex Algorithm Parameterized by the Degeneracy Gap” was reported in the 32nd International Joint Conference on Artificial Intelligence (IJCAI'23) in August 2023 at Macao.}
This work has two main contributions.

\begin{itemize}
	\item We propose an algorithm for the maximum $k$-plex problem with running time $O(|V|^{O(1)}(k+1)^{g_k(G)})$, where $k\ge 1$ is a fixed value and $g_k(G)$ is the gap between the degeneracy bound $d(G)+k$ and the maximum $k$-plex size. When $g_k(G)$ is in $O(\log(|V|))$ (which is often the case empirically), the algorithm runs in polynomial time. As an extension, we also discuss a possibly even smaller parameter $cg_k(G)$, the community-degeneracy gap, and design a corresponding parameterized algorithm. Our main techniques to achieve these results include the degeneracy ordering, the subset enumeration, and a dedicated branching algorithm for the dual \emph{$d$-Bounded-Degree-Deletion} ($d$-BDD) problem.
	
	\item We implement the algorithms and show that they are empirically competitive with the state-of-the-art algorithms for a wide range of instances. In particular, for the group of real-world graphs, our algorithms perform better than existing algorithms with large $k$ values, e.g. $k=15$ and $20$. This is consistent with the fact that the degeneracy gap increases slowly or even decreases as $k$ increases.
	Furthermore, we perform a detailed experimental analysis to illustrate the efficiency and effectiveness of key parameters and algorithmic components, including the degeneracy gap, the branching factor, the graph reduction, and the bound estimation.
\end{itemize}

The paper is organized as follows. In the following section, we present the necessary notations, graph parameters, and existing work for the maximum $k$-plex problem. We also introduce its dual problem, the $d$-BDD problem, which is important for understanding the algorithm.
In Section 3, we illustrate our main algorithm that is parameterized by the degeneracy gap. Then, we extend this algorithm to be parameterized by the community-degeneracy gap. 
In Section 4, we introduce how to implement the algorithms mentioned above. 
Practical speed-up techniques are also studied in this section. 
In Section 5, detailed experiments and analysis of these algorithms are presented. 
Conclusions and perspectives are drawn in the last section. 
Source codes and supplementary materials are available at \url{https://github.com/joey001/kplex_degen_gap}.

\section{Preliminaries}
\subsection{Notations and Problem Definition}
%\subsection{Notations}
Let $G=(V,E)$ be a simple undirected graph with vertex set $V$ and edge set $E$.
The complement graph of $G$ is denoted by $\overline{G} = (V, \overline{E})$, where $\overline{E}=\{(u,v) : v,u \in V, u \neq v, \{u,v\}\notin E\}$.
For any vertex $v$, we use $N_G(v) = \{u\in V: \{u,v\}\in E\}$ to denote the set of neighbors of $v$, and $N^2_G(v) = (\bigcup_{u\in N_G(v)} N_G(u)) \setminus (N_G(v) \cup \{v\})$ to denote the \emph{2-hop neighbors} of $v$, that is, the set of neighbors of the vertices in $N_G(v)$ except $v$ and the vertices in $N_G(v)$ themselves. 
We also use these notations for an edge $e=\{u,v\}$, that is, $N_G(e)= N_G(u)\cap N_G(v)$ denotes the set of common neighbors of $u$ and $v$, and $N^2_G(e)=(N^2_G(u) \cup N^2_G(v)) \setminus (e \cup N_G(e))$ denotes the 2-hop neighbors of both $u$ and $v$ except $\{u,v\}$ and $N_G(e)$ themselves.
Furthermore, given a vertex set $S\subset V$, $G[S]$ denotes the subgraph induced by $S$ and given an edge set $E' \subseteq E$, $G[E']$ denotes the subgraph spanned by $S'$.

Let $k$ be a positive integer, a $k$-plex $P$ is a vertex set such that for any $v\in P$, $| N_G(v) \cap P| \ge |P|-k$.
We denote the size of the maximum k-plex in $G$ as $\omega_k(G)$. 
There are two important properties of $k$-plex. 
First, any subset of a $k$-plex is still a $k$-plex (\citeauthor{trukhanov_algorithms_2013} \citeyear{trukhanov_algorithms_2013}). 
Second, if $P$ is a $k$-plex and $|P| \ge 2k-1$, then $G[P]$ must be a connected graph and the length of the shortest path between two distinct vertices in $G[P]$ is bounded by 2, while a $k$-plex with at most $2k-2$ vertices is probably unconnected (\citeauthor{xiao2017fast} \citeyear{xiao2017fast}, \citeauthor{conte2018d2k} \citeyear{conte2018d2k}). 
In many applications, these trivially small, and probably unconnected $k$-plexes are of no interest. 
Thus, in the paper, we only investigate the problem of finding the maximum $k$-plex of size at least $2k-1$, as in \citep{wang2022listing,chang2022efficient}.

We define the decision version of the maximum $k$-plex problem, namely the $k$-PLEX problem, as follows.

\begin{framed}
\begin{problem}[The $k$-PLEX Problem]
	Given a graph $G=(V, E)$, two positive integers $k$ and $p$ ($p\ge 2k-1$), is there a $k$-plex of size at least $p$ in $G$?
\end{problem}
\end{framed}

\subsection{Important Parameters}
We introduce some important parameters that are used in the work.

\subsubsection{Degeneracy.} 
A graph $G=(V,E)$ is called \emph{$d$-degenerate} if every vertex-induced subgraph has a vertex of degree at most $d$.
Equivalently, a graph $G=(V,E)$ is $d$-degenerate if and only if there exists a $d$-degenerate ordering, $v_1, \dots, v_{|V|}$ such that for each vertex $v_i$, the degree of $v_i$ in the vertex-induced subgraph $G[\{v_i,...,v_{|V|}\}]$ is at most $d$.
The degeneracy parameter of $G$, denoted as $d(G)$, is the smallest value $d$ such that $G$ is $d$-degenerate. 
The corresponding $d(G)$-degenerate ordering is simply called \emph{degeneracy ordering}.
A degeneracy ordering of $G=(V,E)$ can be computed in time $O(|V|+|E|)$ by repeatedly removing a vertex with the minimum degree in the remaining graph until the graph becomes empty \citep{matula1983smallest}. 

\textit{Community-degeneracy} another parameter extended from degeneracy.
A graph $G=(V,E)$ is called \emph{$c$-community-degenerate} if every non-empty edge-induced subgraph $G'$ has an edge $e$ with $|N_{G'}(e)| \le c$. 
It is also equivalent to say that a graph $G=(V,E)$ is $c$-community-degenerate if and only if there exists a $c$-community-degenerate ordering, $e_1, \dots, e_{|E|}$, such that for each edge $e_i$, $|N_{G'}(e_i)|$ in the edge-induced subgraph $G'=G[\{e_i,...,e_{|E|}\}]$ is at most $c$.
Similarly, the community degeneracy parameter of $G$, denoted as $cd(G)$, is the smallest value $c$ such that $G$ is $c$-community-degenerate, and the corresponding $cd(G)$-community-degenerate ordering is simply called \emph{community-degeneracy ordering}.
The computational complexity to compute the community degeneracy parameter is $O(|V||E|)$.  
This can be achieved repeatedly by removing an edge whose incident vertices have the fewest common neighbors \citep{buchanan2014solving}. 

It is known that $cd(G)+1 \le d(G) \le \sqrt{|V|+2|E|}$ \citep{eppstein2011listing}.
But in reality, both the degeneracy and the community-degeneracy are much smaller than the vertex numbers in a sparse graph. 
The community-degeneracy $cd(G)$ is sometimes much smaller than the degeneracy $d(G)$.
For example, the degeneracy of the $d$-dimensional hypercube graph is $d$, while the community-degeneracy of such a graph is $0$ (as it is triangle-free).

For notation convenience, given a degeneracy ordering $v_1,...,v_{|V|}$ of $G=(V,E)$, we use $N_G^+(v_i)$ to denote $N_G(v_i) \cap \{v_{i+1},...,v_{|V|}\}$, and $N_G^{2+}(v_i)$ to denote $N_G^{2}(v_i) \cap \{v_{i+1},...,v_{|V|}\}$;
Given a community-degeneracy ordering $e_1,\dots,e_{|E|}$ of $G=(V,E)$, we use $N_G^+(e_i)$ to denote $N_G(e_i) \cap V'$ and $N_G^{2+}(e_i)$ to denote $N_G^{2}(e_i) \cap V'$, where $V'=\{v \in V : \exists\{v,w\} \in \{e_{i+1},\dots,e_{|E|}\}\}$.

\subsubsection{Degeneracy gaps.} 
For a given graph $G$, we define the parameter \emph{$k$-plex-degeneracy gap} $g_k(G)$ as $d(G)+k-\omega_k(G)$.
Because $d(G)+k$ is an upper bound of $\omega_k(G)$ in a graph $G$ \citep{zhou2017frequency,conte2018d2k}, this parameter is non-negative.
When the context is clear, we simply call the parameter the \emph{degeneracy gap}.
Similarly, the community degeneracy bound $cd(G)+2k$ is also an upper bound of $\omega_k(G)$ \citep{gianinazzi2021parallel,walteros2020maximum,buchanan2014solving}.
We define $cd(G)+2k-\omega_k(G)$ as the \emph{$k$-plex-community-degeneracy gap} $cg_k(G)$ of $G$, which is abbreviated as community-degeneracy gap.

In real-world graphs, the degeneracy gap is usually smaller than the degeneracy.
 For example, when $k=2$, the degeneracy gap of soc-livejournal is only $1$, in contrast to the degeneracy value 213.
It is also noteworthy that as $k$ increases, the degeneracy gap increases slightly or even decreases for most real-world graphs. As an illustration, the degeneracy gap of soc-lastfm with 1,191,805 vertices and 4,519,330 edges is 54 when $k=2$, and drops to 34 when $k=20$.
Interestingly, the community-degeneracy gap can be considerably smaller than the degeneracy gap, especially for small values $k$. 
For example, when $k=2,5$, the community-degeneracy gaps of soc-lastfm are 7 and 4, while the degeneracy gaps are 54 and 48 respectively.

\subsection{Existing Works}
%\paragraph{Theoretical work}
\subsubsection{Theoretical Results}
For the maximum $k$-plex algorithm, the first worst-case running  time guarantee is $O(|V|^{O(1)}\alpha_k^{|V|})$ \citep{xiao2017fast}. 
When $k=1,2,3$ and $4$, $\alpha_k=1.618, 1.839,1.928$ and $1.966$ respectively.
In particular for $k=1$, the running time was improved to  $O(|V|^{O(1)}1.2^{|V|})$ by \cite{xiao2017exact}.
However, this algorithm is mainly of theoretical interest. 

Regarding the parameterized complexity of the problem, \citet{komusiewicz2016multivariate} pointed out that the problem is parameterized by $d(G)$.
\citet{wang2022listing} gave a $O(|V|^{O(1)}\alpha_k^{d(G)})$-time algorithm, where $\alpha_k$ is the same as in \citep{xiao2017fast}.
The degeneracy gap parameter $g(G)$ was first introduced by \citet{walteros2020maximum} for the maximum clique problem. 
They showed that the maximum clique problem has a time complexity $O(|V|^{O(1)}1.28^{g_1(G)})$.
In the thesis of \citet{wunsche2021mind}, they showed that the maximum $k$-plex problem is parameterized by the \textit{$2$-degeneracy gap} $g_2(G)$, a parameter similar to our degeneracy gap.
In brief, the $2$-degeneracy gap is equal to $d_2(G)+1-\omega_k(G)$, where $d_2(G)$ extends $d(G)$ by generalizing the degree of a vertex $v$ to the $2$-degree of $v$, that is, the number of vertices contains all vertices with distance at most $2$ to $v$, except $v$ itself.
In almost all graphs, $g(G)$ is much smaller than $g(G)$. 
A notable work by \cite{figiel2022correlating} discovered that the practical time to solve the maximum $k$-plex is a positive correlate with $d_2(G)$, which also inspires us to investigate the corelation between the practical performance of our algorithm and $g(G)$.

\subsubsection{Practical Results}
There are also several empirical methods without worst-case run-time guarantees. 
\citet{balasundaram2011clique} analysed the linear programming polytope of the maximum $k$-plex problem, which initiated a large body of experimental studies for this problem, including both heuristic and exact algorithms \citep{zhou2017frequency,nogueira2020gpu,pullan2021local, mcclosky2012combinatorial,moser2012exact,trukhanov2013algorithms,shirokikh2013degree,gschwind2018maximum,gao2018exact}.
In recent years, \cite{zhou2021improving} proposed a branch-and-bound algorithm with the so-called first- and second-order reduction and other pruning techniques. \cite{chang2022efficient} revisited their ideas and developed a faster maximum $k$-plex solver, namely kPlexS, for large sparse graphs. 
\begin{comment}
On the other hand, \cite{jiang2021new} studied pruning techniques and implemented an algorithm that is particularly efficient for dense artificial graphs.  
\end{comment}
On the other hand, \cite{jiang2021new} studied bounding and pruning techniques and implemented an algorithm that is particularly efficient for dense artificial graphs.  
To date, the latter two algorithms are believed to be the most competitive in practice.

\subsection{The Dual Problem -- \texorpdfstring{$d$}{d}-BDD Problem}

The $k$-PLEX problem is closely related to the $d$-Bounded-Degree-Deletion ($d$-BDD) problem. A graph $G$ is called \emph{$d$-degree-bounded} if the maximum degree of $G$ is at most $d$. 
\begin{framed}
\begin{problem}[The $d$-BDD Problem]
	Given a graph $G=(V,E)$, two non-negative integer $d$ and $t$, is there a vertex set $D$ of size at most $t$ such that $G\setminus D$ is $d$-degree-bounded?
\end{problem}
\end{framed}

In a graph $G=(V,E)$, there is a $k$-plex of size $p$ if and only if there is a $(k-1)$-bdd of size $|V|-p$ in the complement graph $\overline{G}$.
In this sense, we say that $d$-BDD is the dual problem of $k$-PLEX.
However, the parameterized complexities of the two problems are quite different. 
It is known that $d$-BDD is fixed-parameter tractable (FPT) with respect to parameter $t$, i.e., there exists an algorithm running in time $O(f(t){|V|}^{O(1)})$ where $f$ is a computable function. 
In contrast, $k$-PLEX is W[1]-hard with respect to parameter $p$.
In the literature, \cite{nishimura2005fast} presented a $O((d+t)^{t+3}t+n(d+t))$ algorithm for $d$-BDD, followed by improvements in \citep{moser2012exact,xiao2016dBDD}. 
For $d \ge 3$, the $d$-BDD problem can be solved in $O(|V|^{O(1)}(d+1)^t)$ by \cite{xiao2016dBDD}. However, these algorithms are only of theoretical interest at this stage. In our algorithm, we will adopt a simple and easy-to-implement $d$-BDD algorithm as a subroutine.

We further emphasize that it is not computationally viable to apply these FPT algorithms of $d$-BDD with $\overline{G}$ directly.
In large real-world graphs, the maximum $k$-plex size is often very small, while the vertex number is quite large.
As a result, the parameter $t=|V|-p$ could be quite large, making the above FPT algorithms of $d$-BDD inefficient in practice.
\cite{moser2012exact} have tried to solve the maximum $k$-plex problem in this way, but their algorithm is somehow better suited to dense artificial graphs than to large real-world graphs.

\section{Our Algorithm Frameworks}
\subsection{The Algorithm Parameterized by Degeneracy Gap}
We first present KPLEX in Alg. \ref{Alg_framework} for solving the $k$-PLEX problem.
Notably, the algorithm is parameterized by $g$, where $g=d(G)+k-p$. We justify its correctness as follows.

\begin{algorithm}[ht!]
	\caption{KPLEX parameterized by $g=d(G)+k-p$}
	\label{Alg_framework}
	\emph{KPLEX}$(G,k, p)$\\
	\KwIn{An input graph $G=(V,E)$, two positive integers $k$ and $p\ge 2k-1$.}
	\KwOut{A $k$-plex of size at least $p$ or 'NO' if there is no such set in $G$.}
	\Begin{
  		Sort $V$ by degeneracy ordering $v_1, ..., v_{|V|}$.\\
		\For {$v_i$ from $v_1$ to $v_{|V|}$ 
		} {
			\For {$S\subseteq N^{2+}_{G}(v_i)$ and $|S|\le k-1$}{ 
				$G_s=(V_s, E_s) \gets G[\{v_i\} \cup S \cup N^+_{G}(v_i)]$\\
				$d \gets k-1$, $t \gets |V_s|-p$  \\
				$D^*\gets$ DBDD($\overline{G_s}$, $d$, $t$, $N_{G}^+(v_i)$, $\emptyset$)\\
				\If{$D^*\neq$ 'No' }{				
					\Return $V_s \setminus D^*$
				}
			}
		}
        \Return {'No'}
	}
	
\end{algorithm}

 Because the distance between any two vertices in a $k$-plex $P$ ($|P| \ge 2k-1$) is at most 2 in $G[P]$, the following observation holds.
\begin{observation}
	\label{observation_1}
	In a graph $G=(V,E)$, let $P\subseteq V$ be an arbitrary $k$-plex such that $|P|\ge 2k-1$. Denote $v_i$ as the first vertex in $P$ with respect to the degeneracy ordering of $G$. 
	Then $P$ can be split into three subsets $\{v_i\}$, $N^+_G(v_i)\cap P$ and $N^{2+}_G(v_i)\cap P$, where $|N^{2+}_G(v_i)\cap P|\le k-1$.
\end{observation}

By the above observation, we design the algorithm KPLEX using the following idea. 
For each $v_i$, we enumerate all possible subsets $S\subseteq N^{2+}_G(v_i)$ satisfying $|S|\le k-1$.
%$S\subseteq N^{2+}_G(v_i)$.
%Because $|N^{2+}_G(v_i)\cap P|\le k-1$,  we enumerate
For one vertex $v_i$ and one subset $S$, we decide if there is a $k$-plex of size at least $p$ that includes $\{v_i\}\cup S$ in $G[\{v_i\} \cup S\cup N^+_G(v_i)]$.
If so, then we find one $k$-plex of size at least $p$. 

%KPLEX generally implements the above idea. 
To be more specific, for a vertex $v_i$ from $v_1$ to $v_{|V|}$ and a subset $S \subseteq N^{2+}_G(v_i)$, KPLEX induces a subgraph $G[\{v_i\}\cup S\cup N^+_G(v_i)]$ as $G_s=(V_s,E_s)$.
% (In fact, the subgraph $G_s$ is further reduced by reduction rules proposed in the implementation part). 
In lines 6-10, KPLEX decides if there is a $k$-plex of size  at least $p$ that includes $\{v_i\} \cup S$ in $G_s$.
However, in line 8, instead of solving this decision problem directly, KPLEX solves the dual problem -- whether there is a $(k-1)$-bdd of size at most $|V_s|-p$ from $N^+_G(v_i)$ in graph $\overline{G_s}$, by invoking a subroutine named DBDD. The description of DBDD is given in Alg. \ref{alg_bdd}.
% Moreover, this subroutine will be further enhanced with bounding technique mentioned in the implementation part.

\begin{algorithm}[htbp!]
    % \small
	\caption{The $d$-BDD algorithm \label{alg_bdd}}
	\emph{DBDD}$(G, d, t, C, D)$\\
	\KwIn{A graph $G=(V,E)$, two non-negative integers $d$ and $t$, a candidate vertex set $C$, a growing vertex set $D$ that $D\cap V=\emptyset$.}
	\KwOut{A $d$-bdd $D$ of size at most $t$ or 'No' if there is no such vertex set.}
	\Begin{
            % $t$, $C$, $D$ $\gets$ Update$(G, d, t, C, D)$ \\	
		\If{$ t < 0 $ or $\exists u \in V \setminus C$ that $|N_{G}(u)\setminus C| > d$}{
               \Return{'No'}
		}            
        \If{$\forall u\in V, |N_{G}(u)|\le d$}{
               \Return {$D$}
		}
		\If{$\exists u\in C$ that $N_G(u) > d+t$}{ %u must in B
			\Return {DBDD$(G \setminus \{u\}, d, t-1, C\setminus \{u\}, D \cup \{u\})$}
		}
% 		\If{$\exists u\in C$ that $N_G(u) > d+t$}{ %u must in B
% 			\Return {DBDD$(G \setminus \{u\}, d, t-1, C\setminus \{u\}, D \cup \{u\})$}
% 		}
		\If{$\exists u \in C$ that $\forall v\in \{u\}\cup N_G(u), N_G(v)\le d$}{ %u must not in B, and can be removed
			\Return{DBDD$(G,d, t, C\setminus \{u\}, D)$}
		}
        Pick a vertex $u_p\in V$ of maximum degree in $G$.\\
        Suppose $N_G(u_p)\cap C$ as $v_1,...,v_{s}$ in arbitrary ordering, where $s \gets |N_G(u_p)\cap C|$. \\
        \If{$u_p \in C$}{
            $b\gets d+1-|N_G(u_p)\setminus C|$\\
            (Br. $1$) 
		$D_1\gets$ DBDD$(G\setminus\{u_p\}, d, t-1, C\setminus\{u_p\}, D\cup\{u_p\})$ \\
            (Br. $i\in \{2,...,b\}$) 
		$D_{i}\gets$ DBDD$(G\setminus\{v_{i-1}\}, d, t-1, C\setminus\{u_p,v_1,...,v_{i-1}\}, D\cup\{v_{i-1}\})$ \\
		(Br. $(b+1)$) 
		$D_{b+1}\gets$ DBDD$(G\setminus\{v_{b},...,v_s\}, d, t-1-s+b, C\setminus \{u_p,v_1,...,v_s\}, D\cup \{v_{b},...v_{s}\})$\\
        }
        \Else{
            $b\gets d-|N_G(u_p)\setminus C|$\\
            (Br. $i\in \{1,...,b\}$) 
		$D_{i}\gets$ DBDD$(G\setminus\{v_i\}, d, t-1, C\setminus\{v_1,...,v_i\}, D\cup\{v_i\})$ \\
		(Br. $(b+1)$) 
		$D_{b+1}\gets$ DBDD$(G\setminus\{v_{b+1},...,v_s\}, d, t-s+b, C\setminus \{v_1,...,v_s\}, D\cup \{v_{b+1},...v_{s}\})$\\
        }
        %sort $N_G[u_p]\cap C$ in descending ordering $v_1,...,v_{s}$.\\
        
		%let $X={v_1, v_2,...,v_{|X|}}$ where $v_1=u_p$ and $v_2,...,v_{|X|}$ are vertices of $N_G(u_p)\cap C$ in arbitrary order\\
		\If{$\exists D_i\neq$ 'No'}{
			\Return {$D_i$}
		}
            \Return {'No'}
	}
\end{algorithm}

%{\color{red}First, we show the BDD algorithm (Alg. \ref{alg_bdd}) solves the so-called $C$-candidate $d$-BDD problem.}
%without giving a tedious formal proof.
In terms of the input of DBDD, $D$ is a set of growing vertices, i.e., a set that  must be a part of the $d$-bdd solution. So $D$ is empty initially. $C\subseteq V$ is the candidate set, i.e., the target $d$-bdd set, if exists, must be a subset of $C$.
It is clear that DBDD is a tree search algorithm. 
At each node (or each invocation), DBDD first reduces the input size or decides the solution directly.
When the input cannot be solved or reduced anymore, it branches, i.e., DBDD calls itself multiple times with different inputs that cover all possibilities, see \cite{cygan2015parameterized}.
%Reductions reduce the input into a smaller one or stop the search while branchings 
For example, if a vertex $u\in C$ is selected as a member of $D$, then $u$ should be removed from $G$ and $C$, and meanwhile $t$ is reduced by 1 in the recursive call; if a vertex $u \in C$ is excluded from being a member of $D$, $u$ should be removed from $C$ without changing $t$.  
Now, we have Lemma \ref{correctness_bdd} holds.

\begin{lemma}
\label{correctness_bdd}
Given $G=(V,E)$, a candidate set $C\subseteq V$ and an integer $t\le |C|$, DBDD($G,d, t, C, D$) correctly finds a $d$-bdd set $D^*\subseteq C$ such that $|D^*| \le t$ or return 'No' if no such set exists. 
\end{lemma}

\proof{Proof.}
% \begin{proof}
%Note that from lines 3-10, we do reductions while for the rest lines, we do branching.
Given the input $G, d, t, C$ and a growing set $D$ for DBDD, we have the following reduction rules.

\begin{enumerate}    
    \item If $t<0$ or $G \setminus C$ is not $d$-degree-bounded, then there is no $d$-bdd set of size at most $t$ in the current input. 
    \item If $\exists u\in C$ that $N_G(u) > d+t$, then $u$ must be in any $d$-bdd set of size at most $t$.
    \item If $\exists u\in C$  that $\forall v\in \{u\} \cup N_G(u), N_G(v)\le d$, then $u$ must be excluded from some $d$-bdd set of size at most $t$.
\end{enumerate}

The first rule holds straightforwardly, and the second and third are from \cite{moser2012exact}.
These reduction rules are implemented in lines 3-10 in Alg. \ref{alg_bdd}.
%Then, if $t=0$ and $G$ is $d$-degree-bounded, then $D$ is a $d$-BDD set and we also stop the search and return $D$ as the solution $D^*$.
%As for other two cases that $\exists u\in C$, $N_G(u) > d+t$ and $\forall v\in \{u\} \cup N_G(u), N_G(v)\le d$, $u$ must be and must not be a vertex in $D$, respectively. (See \cite{moser2012exact} for the correctness). In either of the cases, we removed $u$ from candidate set $C$.
When the above reduction rules cannot be applied any more, the current input is in a state that $G$ is not $d$-degree-bounded and $t>0$, so there must exist a vertex $u_p \in V$ that $|N_G(u_p)|> d$.
By definition, if there is a solution in the current input, then either $u_p$ is in the solution or at least $|N_G(u_p)|-d$ vertices from $N_G(u_p)$ are in the solution.
Assume $D^*\subseteq C$ is a solution with $|D^*| \le t$ in the current input.
For illustration purpose, denote $N_G(u_p)\cap C=\{v_1,...,v_{s}\}$ in arbitrary ordering, $s$ being the size of $N_G(u_p)\cap C$. It is easy to check that the following cases are disjoint and complete.
\begin{itemize}
    \item $u_p \in C$. There are $b+1$ possibilities where $b=d+1-|N_G(u_p)\setminus C|$.
     \begin{enumerate}
         \item The first possibility, $u_p\in D^*$.
         \item The $i$-th possibility where $i\in \{2,...,b\}$, $u_p,v_1...,v_{i-2}\notin D^*$ and $v_{i-1}\in D^*$. (In case $i=2$, $u_p\notin D^*$ and $v_1\in D^*$.)
         \item The $b$+1-th possibility, $u_p,v_1,...,v_{b-1} \notin D^*$ and $v_{b},...,v_s \in D^*$.
     \end{enumerate}
    \item $u_p \notin C$. There are $b+1$ possibilities where $b=d-|N_G(u_p)\setminus C|$.
    \begin{enumerate}
        \item The $i$-th possibility where $i\in \{1,...,b\}$, $v_1,...,v_{i-1}\notin D^*$ and $v_i\in D^*$. (In case $i=1$, $v_1\in D^*$.)
        \item The $b$+1-th possibility, $v_{1},...,v_{b}\notin D^*$ and $v_{b+1},...,v_s\in D^*$.
    \end{enumerate}
\end{itemize}
%With simple calculation, one can check that the above cases covers all possibilities. 

Since DBDD exactly covers all the above cases (lines 11-21), we conclude that DBDD is  complete and correct.
% \end{proof}
\Halmos
\endproof

%Now, we are in a state that $G$ is not $d$-degree-bounded but we do not know which vertices can be selected into $D$ such that the remaining graph is $d$-degree-bounded. We branch to enumerate all possibilities.
%First, if $u\in D^*$ and $|N_G(u)|\le d$, we can safely remove $u$ from $D^*$ without violating that $D^*$ is still a $d$-BDD. 
%As $G$ is not $d$-degree-bounded, there must exist a vertex $u_p \in V$ that $|N_G(u_p)|> d$.
%By definition, we can conclude that either $u_p$ is in $D$ or at least $|N_G(u_p)|-d$ vertices are in $D$.
%However, vertices of $N_G\setminus C$ are not in $B^*$ by problem definition. 
%By definition, we can conclude that either $u_p$ is in $D^*$ or 
%Thus, from lines 13 to 21 in the algorithm, we implement the branching idea which covers all the possibilities.
%If $u_p \in C$, we select $u_p$ into the growing set $D$ in the first branch, and exclude it from being a member of $D$ in the remaining branches. 
%Note that, in the last branch, we exclude $u_p$ and $d$ of its neighbors from $D$, indicating that the remaining part of vertices in $N_G(u_p)\cap C$ must be a part of $D$. In total, we generate $d+2-|N_G(u_p)\setminus C|$ branches.
%For the other case where $u_p \notin C$, we do not need to consider the possibility that $u_p$ is a member of $D$. Hence, the branching factor is one less than the former one.
%To sum up, the algorithm is complete -- It finds a solution or report 'No' if no such solution exists.

\subsubsection{Running Time Analysis}
\paragraph{Running time of DBDD.}
As we mentioned, DBDD is a tree search algorithm. We can measure its worst-case running time by simply multiplying the number of tree nodes with the complexity of producing a tree node.
%\textcolor{red}{[Delete], as suggested by the monograph of \cite{cygan2015parameterized}.}
Suppose the input of the our algorithm is DBDD($G=(V,E),d,t,C,D$). 
We can safely assume that the time taken at each tree node is $O(|V|^{O(1)})$. (In our implementation, the time is bounded by $O(|V|^2)$.) 
%Also, we use $T(t)$ to use the number of tree nodes given input BDD($G=(V,E),d,t,C,B$).
By the branching rules, the parameter $t$ decreases at least 1 at each sub-node.
%Thus, given the current node DBDD($G=(V,E),d,t,C,D$), 
Denote $L(t)$ as the number of leaf nodes in the subtree.  
We have
\[
L(t) \le \underbrace{L(t-1)+...+L(t-1)}_{b \text{ times}}+L(t-N_G(u_p)+d),
\]
where $b=d+1-|N_G(u_p)\setminus C|$ when $u_p\in C$ and $b=d-|N_G(u_p)\setminus C|$ when $u_p\notin C$, and $L(1)=1$.
In the worst case, $u_p\in C$, $N_G(u_p)\setminus C=\emptyset$ and $N_G(u_p)=d+1$, that is to say, $b=d+1$ and $N_G(u_p)-d=1$. 
So, we obtain $L(t) \le (d+2)^{t}$ and the number of all tree nodes is $O((d+2)^{t})$.
Combining the running time of each node, we conclude the running time $T_{DBDD}(G,d,t,C,D)$ is bounded by $O(|V|^2(d+2)^{t})$.

\paragraph{Running time of KPLEX.}
Suppose the input of  KPLEX as $(G=(V,E),k,p)$, the running time of KPLEX $T_{KPLEX}(G,k,p)$ can be calculated by
\begin{small}
\begin{align*} 	 
 	&T_{KPLEX}(G,k,p) = T_{deg}(G)+ \\
 	&\sum_{i=1}^{|V|}\sum_{\substack{S\subseteq N^{2+}_G(v_i) \\ |S|\le k-1}}  
 	{\Bigl(T_{graph}(G_s)+T_{DBDD}(\overline{G_s}, k-1,t, N_G^+(v_i), \emptyset)\Bigr)},
\end{align*}
\end{small}
where $T_{deg}(G)$ is the time of degeneracy ordering, $T_{graph}(G_s)$ is the time of building $G_s$ given a vertex $v_i$ and a subset $S$, and $T_{DBDD}(\overline{G_s}, k-1, t, N_G^+(v_i), \emptyset)$ is the time of running DBDD$(\overline{G_s}, k-1, t, N_G^+(v_i), \emptyset)$.
Note that $G_s=(V_s,E_s)$ is the vertex-induced subgraph $G[\{v_i\}\cup S \cup N^+_{G}(v_i)]$ and $t=|V_s|-p$.

It is known that $T_{deg}(G)\le O(|V|+|E|)$ by \cite{batagelj2003m}.
Define $g=d(G)+k-p$. Because $|N^+_{G}(v_i)|\le d(G)$ and $|S|+1\le k$, we have $|V_s| \le d(G)+|S|+1 \le d(G)+k$ and $t \le d(G)+k-p = g$.
Therefore, $T_{graph}(G_s)\le (d(G)+k)^2$  and $T_{DBDD}(\overline{G_s}, k-1, t, N_G^+(v_i), \emptyset)\le (d(G)+k)^2(k+1)^{g}$. (The first inequality is obtained by using adjacency matrix to build a graph and the second inequality is obtained by the above analysis of DBDD.)
Therefore,
\begin{small}
	\begin{align*} 	 
		&T_{KPLEX}(G,k,p) = O\biggl(|V|+|E|+ \\
		&\sum_{i=1}^{|V|}\sum_{\substack{S\subseteq N^{2+}_G(v_i) \\ |S|\le k-1}}  
		{\Bigl((d(G)+k)^2+(d(G)+k)^2(k+1)^{g}\Bigr)}\biggr) \\
		&=O\biggl(|V|+|E|+\sum_{i=1}^{|V|}{|V|^{k-1}(d(G)+k)^2(k+1)^{g}}\biggr) 	\\
		&=O(|V|+|E|+|V|^k(d(G)+k)^2(k+1)^{g}).		
	\end{align*}
\end{small} 
In summary, we obtained the running time for KPLEX as follows.

\begin{lemma}
	\label{lemma-KPLEX-complex} 
	Given a graph $G=(V, E)$, a fixed integer $k\ge 1$ and an integer $p$ that $p \ge 2k-1$, KPLEX$(G,k,p)$ solves the $k$-PLEX problem in time $O(|V|^{O(1)}(k+1)^g)$, where $g=d(G)+k-p$.
\end{lemma}

Note that the above analysis justified Lemma \ref{lemma-KPLEX-complex}. With this conclusion, it is easy to use KPLEX for solving the maximum $k$-plex problem:  For each integer $p$ from $d(G)+k$ to $2k-1$, we call KPLEX$(G,k,p)$ to decide if there exists a $k$-plex of size at least $p$ in $G$. If so, then we stop and conclude that $\omega_k(G)$ is equal to $p$.  If no for any $p$ value, then we conclude that $\omega_k(G)<2k-1$, i.e., the maximum $k$-plex is trivially small. To sum up, we run KPLEX at most $|V|$ times, and at each time, $g\le d(G)+k-\omega_k(G)$. Hence, the following conclusion holds.

\begin{theorem}
    Given a graph $G=(V, E)$ and a fixed integer $k\ge 1$, the maximum $k$-plex problem can be solved in time $O(|V|^{O(1)}(k+1)^{g_k(G)})$ with KPLEX, where $g_k(G)=d(G)+k-\omega_{k}(G)$.
\end{theorem}

\noindent \textbf{Remarks} We observed that in real-world graphs, $g_k(G)$ is often small. 
If we assume $g_k(G)$ is bounded by $O(\log{|V|}))$ (which is often the case), we can solve the maximum $k$-plex problem in polynomial time.
Moreover, we observed that the worst-case branching factor of our tree search, $k+1$, is very pessimistic in practice. 
For example, for the consph graph with $k=20$, the branching factors in average is as small as 1.84 in the experiments.

\subsection{Extension of KPLEX for Community-Degeneracy Gap}

To design a maximum $k$-plex algorithm parameterized by community-degeneracy gap, it is natural to extend the KPLEX algorithm such that it is parameterized by another gap-like parameter, i.e. $cg=cd(G)+2k-p$. In Alg. \ref{Alg_framework2}, we show the extended algorithm KPLEX$\mathrm{_{com}}$.

\begin{algorithm}[ht!]
	%\scriptsize
	\caption{KPLEX$\mathrm{_{com}}$ parameterized by $cg=cd(G)+2k-p$}
	\label{Alg_framework2}
	\emph{KPLEX$\mathrm{_{com}}$}$(G,k, p)$\\
	\KwIn{An input graph $G=(V,E)$, two positive integers $k$ and $p\ge 2k-1$.}
	\KwOut{A $k$-plex of size at least $p$ or 'NO' if there is no such set in $G$.}
	\Begin{
  		Sort $E$ by community-degeneracy ordering $e_1, ..., e_{|E|}$.\\			
		\For {$e_i$ from $e_1$ to $e_{|E|}$ 
		} {
			\For {$S\subseteq N^{2+}_{G}(e_i)$ and $|S|\le 2k-2$}{ 
				%\tcc{$S$ is a trivial $k$-plex}	
				$G_s=(V_s, E_s) \gets G[\{u, v\} \cup S \cup N^+_{G}(e_i)]$  \tcp{$e_i=\{u,v\}$}
				$d \gets k-1$, $t \gets |V_s|-p$  \\
				$D^*\gets$ DBDD($\overline{G_s}$, $d$, $t$, $N_{G}^+(e_i)$, $\emptyset$)\\
				\If{$D^*\neq$ 'No' }{				
					\Return $V_s \setminus D^*$
				}
			}
		}
        \Return {'No'}
	}
\end{algorithm}

Specifically, in lines 4-5, KPLEX$\mathrm{_{com}}$ first enumerates each edge $e_i=\{u,v\}$ from $e_1$ to $e_{|E|}$, and then enumerates each subset $S \subseteq N^{2+}_G(e_i)$ where $|S| \leq 2k-2$. Let us denote $G[\{u,v\} \cup S \cup N^+_G(e_i)]$ as $G_s=(V_s,E_s)$. 
In lines 6-10, KPLEX$\mathrm{_{com}}$ decides if there is a $k$-plex of size at least $p$ that includes $\{u, v\} \cup S$ in $G_s$.
In line 8, KPLEX$\mathrm{_{com}}$ calls the DBDD subroutine to solve the dual problem  -- whether there is a $(k-1)$-bdd of size at most $|V_s|-p$ from $N^+_G(e_i)$ in graph $\overline{G_s}$.

The time complexity of KPLEX$\mathrm{_{com}}$ can be analyzed in a similar way for the KPLEX algorithm. We leave the detailed proof in the Appendix. Finally, we can give the following result.
\begin{theorem}
	Given a graph $G=(V, E)$ and a fixed integer $k\ge 1$, the maximum $k$-plex problem can be solved in time $O(|V|^{O(1)}(k+1)^{cg_k(G)})$ with KPLEX$\mathrm{_{com}}$, where $cg_k(G)=cd(G)+2k-\omega_{k}(G)$. \label{thm-commu}
\end{theorem}

\textbf{Remark.} Both KPLEX and KPLEX$\mathrm{_{com}}$ can be adapted to enumerate large maximal $k$ -plexes, that is, maximal $k$ -plexes with size at least $p$, with time complexity $O(|V|^{O(1)}(k+1)^{g})$ and $O(|V|^{O(1)}(k+1)^{cg})$ respectively, where $g=d(G)+k-p$ and $cg=cd(G)+2k-p$.

\section{Practical Implementation Techniques}
In this section, we introduce implementation techniques for solving the maximum $k$-plex problem in practice. 
We present the whole algorithm Maple in Alg. \ref{alg_max}.
Maple implements the idea of KPLEX, but incorporates several notable features.
First, we start from a lower bound $l$ found by a simple greedy heuristic in \citep{zhou2021improving}. 
This heuristic also relies on a degeneracy ordering of the vertices in the input graph.
Maple uses several speedup techniques to make the algorithm more practical.
For example, in line 8, vertex $v_i$ that is unpromising for a larger $k$-plex is omitted.
In line 11, the graph $G_s$ is reduced but we still ensure that there is at least one $k$-plex of size $p$ in the reduced graph $G_s$ if and only if there is one in the original graph. 
Another change is incorporated in the implementation of the DBDD subroutine----we use some bounding techniques to prune the tree search.
These speedup rules are further illustrated in the following sections.

\begin{algorithm}[htbp!]
	\caption{Maple parameterized by $g_k(G)=d(G)+k-\omega_k(G)$}
	\label{alg_max}
	\emph{Maple}$(G,k)$\\
	\KwIn{An input graph $G=(V,E)$, a positive integer $k$.}
	\KwOut{The maximum $k$-plex of size at least $2k-1$ or 'No' if there is no such set in $G$.}
	\Begin{
		Compute a heuristic solution $P^*$ and a lower bound $l=|P^*|$ \\ via sorting $V$ by degeneracy ordering $v_1,...,v_{|V|}$. \\
		\If{ $l \le 2k-2$}{ 
			$P^* \gets $ \textit{'No'}, $l \gets 2k-2$
		}
		\For {$p$ from $d(G)+k$ to $l+1$ } {
			\For {$v_i$ from $v_1$ to $v_{|V|}$ where $|N_G^+(v_i)|+k\ge p$} {
				\For {$S\subseteq N^{2+}_{G}(v_i)$ and $|S|\le k-1$}{ 
					$G_s=(V_s, E_s) \gets G[\{v_i\} \cup S \cup N^+_{G}(v_i)]$\\
					$G_s=$Reduce$(G_s,v_i, S, k,p)$ \tcp{Graph Reduction}
					$d \gets k-1$, $t \gets |V_s|-p$  \\
					$D^*\gets$ DBDD($\overline{G_s}$, $d$, $t$, $N_{G}^+(v_i)$, $\emptyset$) \tcp{Pruning the search with more bounding technique in DBDD} %($G,d,C$) use in lines 10 and 11 
					\If{$D^*\neq$ 'No' }{				
						\Return $V_s \setminus D^*$
					}
				}
			}
		}
		\Return $P^*$
	}
\end{algorithm}

\subsection{Graph Reduction}
The subgraph $G_s$ is to be built for each $v_i$ and $S\subseteq N^{2+}_{G}(v_i)$ in KPLEX. 
Here, we introduce some reduction rules to reduce the size of $G_s$ or even to determine that the computation of $G_s$ is unnecessary.

\begin{reduce}[First- and Second-order Reduction by \cite{zhou2021improving}]
	\label{common_2}
	Given a graph $G=(V, E)$ and a vertex $u \in V$, if $|N_G(u)| < p-k$, then $u$ is not in any $k$-plex of size at least $p$. Furthermore, for any two distinct vertices $u$ and $v$ of $G$, 
	\begin{enumerate}
		\item if $\{u, v\} \in E$ and  $|N_G(\{u, v\})| < p-2k$, then $u$ and $v$ are not in any $k$-plex of size at least $p$ at the same time.
		\item if $\{u, v\} \notin E$ and $|N_G(\{u, v\})| < p-2k+2$, then $u$ and $v$ are not in any $k$-plex of size at least $p$ at the same time.
	\end{enumerate}
\end{reduce}

The first-order reduction was used in many existing algorithms \citep{zhou2017frequency,gao2018exact,jiang2021new}.
The second-order reduction first appeared in  \citep{zhou2021improving} and was later used in the algorithms presented in \citep{jiang2021new,chang2022efficient}.
We extend these reduction rules to the following \emph{higher-order reduction} rule. 
\begin{reduce}[Higher-order Reduction]
	\label{common_n}
	Given a graph $G=(V, E)$ and an arbitrary vertex set $P\subseteq V$, denoting $n=|P|$ and $\lambda=|E(G[P])|$, if  $|N_G(P)| < p-nk+n(n-1)-2\lambda$, then $P$ is not a subset of any $k$-plex of size at least $p$.
\end{reduce}

Using the higher-order reduction rule, more vertices and edges can be reduced than by using only the first- and second-order reduction rules.
We present the whole reduction procedure in Alg. \ref{alg_reduce}. 
For each vertex $v_i$ and subset $S$, let $P=\{v_i\} \cup S$, $n=|S|+1$ and $\lambda = |E(G_s[\{v_i\} \cup S])|$. With the higher-order reduction, we can save the invocation of DBDD when  $|N_{G_s}(\{v_i\} \cup S)| < p-nk+n(n-1)-2\lambda$. If the invocation to DBDD is unavoidable, the size of $G_s$ can still be reduced by this rule. 

\begin{algorithm}[htbp!]
	\caption{Graph reduction for KPLEX}
	\label{alg_reduce}
	Reduce$(G_s=G[\{v_i\} \cup S \cup N_G^+(v_i)],v_i, S, k,p)$\\
	\Begin{
			\Repeat{No vertex and edge can be removed} {
				\If { $\exists$ $u \in V_s$ s.t. $|N_{G_s}(u)|< p-k$} {
				% 	$V_s\gets V_s\setminus \{u\}$ \\
                    Remove $u$ and its incident edges.
				}
				\If { $\exists$ $u, v \in V_s$ s.t. $u\in N_{G_s}(v)$ and $|N_{G_s}(\{u, v\})| < p-2k$} {
                    Remove edge $\{u, v\}$ from $G_s$.
				}
			}
			$Hop1 \equiv N_{G_s}(v_i)$, $Hop2 \equiv N_{G_s}^2(v_i)$ \\
			\Repeat{No vertex and edge can be removed} {
				\If {$\exists$ $u\in Hop1$ s.t. $|N_{G_s}(\{v_i, u\})|< p-2k$} {
					Remove $u$ and its incident edges.
				}
				\If {$\exists$ $u, v\in Hop1$ s.t. $u\in N_{G_s}(v)$ and $|N_{G_s}(\{v_i, u, v\})| < p-3k$} {
					Remove edge $\{u, v\}$ from $G_s$.
				}
			}
			\Repeat{No vertex and edge can be removed} {
				\If {$\exists$ $u\in Hop2$ s.t. $|N_{G_s}(\{v_i, u\})|< p-2k+2$} {
					% $Hop2\gets Hop2\setminus \{u\}$ \\
					Remove $u$ and its incident edges.
				}
				\If {$\exists$ $u\in Hop1$, $v\in Hop2$ s.t. $u\in N_{G_s}(v)$ and $|N_{G_s}(\{v_i, u, v\})| < p-3k+2$} {
                    Remove edge $\{u, v\}$ from $G_s$.
				}
				\If {$\exists$ $u, v\in Hop2$ s.t. $u\in N_{G_s}(v)$ and $|N_{G_s}(\{v_i, u, v\})| < p-3k+4$} {
					Remove edge $\{u, v\}$ from $G_s$.
				}
			}
      $n \gets |S|+1$, $\lambda \gets |E(G_s[\{v_i\} \cup S])|$ \\
      \If{$\{v_i\} \cup S \nsubseteq V_s$ or $|N_{G_s}(\{v_i\} \cup S)| < p-nk+n(n-1)-2\lambda$}{
        \Return{$G[\emptyset]$}
      }
      \Return{$G_s$}
    }
\end{algorithm}

\subsection{Bound Estimation}

In this section, we introduce some bounding techniques that are used in Maple.

\subsubsection{Initial Lower Bound}

The size of any $k$-plex is a feasible lower bound of the optimal. 
In our implementation, an initial $k$-plex is identified in conjunction with the degeneracy ordering. This is shown on line 3 of Alg. \ref{alg_max}.
Recall that a degeneracy ordering is obtained by repeatedly removing a vertex of the minimum degree.  
So, we keep track of the set of remaining vertices. And when it forms a $k$-plex, we use it as our lower bound solution $P^*$. Therefore, this lower bound is computed in linear time. We also notice that this simple lower bound is usually close to $\omega_k(G)$. For example, when $k=2,5,15$,  the lower bound of the soc-livejournal graph obtained by this heuristic is equal to the optimal.

\subsubsection{Upper Bound of \texorpdfstring{$G_s$}{Gs}}
We use a simple degree bound to estimate the size of $G_s$ in the algorithm. This is shown in line 8 of Alg. \ref{alg_max}.
Given a degeneracy ordering $v_1,...,v_n$, we call $|N_G^+(v_i)|+k$ the \textit{degree bound} of $v_i$ with respect to this degeneracy ordering. 
That is, the number of vertices in graph $G_s=(V_s,E_s)$ can be no more than $|N_G^+(v_i)|+k$.
Therefore, we use this upper bound to avoid unnecessary searches within $G_s$.

\subsubsection{Partition Bound for Minimum \texorpdfstring{$d$}{d}-BDD.}

Let us first formulate the $d$-bdd bounding problem -- What is the minimum number of vertices that must be removed from a set of candidates $C\subseteq V$ in a graph $G=(V,E)$ in order to make the remaining graph $d$-degree-bounded?
Denote $V \setminus C$ as $S$ and assume $S=\{v_1,v_2,\cdots,v_q\}$ where $q = |S|$.
The partition bound is defined as follows.

Define a partition $\Pi=\{\pi_0,\pi_1,\cdots,\pi_q\}$ of $C$ that satisfies the following conditions:
(i) $\bigcup_{i=0}^{q}\pi_i=C$ and $\pi_i \cap \pi_j = \emptyset$ ($0 \le i,j \le q$, $i \neq j$), (ii) $\pi_0=C \setminus \bigcup_{i=1}^{q} N_G(v_i)$, (iii) for $1 \le i \le q$, $\pi_i \subseteq C \cap N_G(v_i)$. Define the array $\Delta = (\delta_1, \delta_2, \cdots , \delta_q)$ of $S$, where $\delta_i = | S \cap N_G(v_i)|$. Then, the following statement holds. 

\begin{lemma}
    \label{lemma_bound}
    With a partition $\Pi$ of $C$ and the $\Delta$ array of $S$, the lower bound of the minimum $d$-bdd excluding $S$ in $G$ can be computed as $|C|-|\pi_0|-\Sigma_{i=1}^{q}min(d-\delta_i,|\pi_i|)$.
\end{lemma}
Indeed, the partition bound was originally proposed by \citep{jiang2021new} for upper bounding the maximum $k$-plex size. 
The pseudo-code for computing such a bound for minimum $d$-bdd size is given in Alg. \ref{alg_bound}.
Here, we use the bound to prune the DBDD tree search algorithm, which is called in line 13 in Alg. \ref{alg_max}. If this bound is greater than $t$, we skip the current tree node without further branching.

\begin{algorithm}[htbp!]
	%\scriptsize
	\caption{Partition Bound for DBDD}
	\label{alg_bound}
	\emph{PartitionBound}$(G,d,C)$\\
	\KwIn{An input graph $G=(V,E)$, an non-negative integer $d$ and a candidate set $C$.}
	\KwOut{The lower bound of minimum $d$-bdd size from $C$ in $G$.}
	\Begin{
	    $S \gets V \setminus C$, $q \gets |S|$, $C' \gets C$\\
	    Sort $S$ as $v_1,...,v_q$ in decreasing order of $| S \cap N_G(v_i)|$\\
		\For {$v_i$ from $v_1$ to $v_q$ } {
		    $\delta_i \gets | S \cap N_G(v_i)|$\\
		    $\pi_i \gets C' \cap N_G(v_i)$, $C' \gets C' \setminus N_G(v_i)$
		}
		$\pi_0 \gets C'$\\
		\Return $|C| - |\pi_0| -\Sigma_{i=1}^{q} min\{d - \delta_i, |\pi_i|\}$\\
	}
\end{algorithm}

\section{Experiments}
In this section, we evaluate our algorithms empirically.  
Our algorithms are written in C++11 and compiled by G++ version 9.3.0 with -Ofast flag. All experiments are conducted on a machine with an Intel(R) Xeon(R) Gold 6130 CPU @ 2.1GHz and an Ubuntu 22.04 operating system. Hyperthreading and turbo techniques are disabled for a steady clock frequency.

As far as we know, the existing algorithms are only tested with $k$ values up to  $7$. 
But the performance at even higher $k$ values should also be an important metric for maximum $k$-plex algorithms.
Therefore, we run experiments with $k=2,5,10,15,20$ using a time limit of 1800 seconds. Our experiments consist of two parts, i.e., an overall performance evaluation and an analysis of key components.

\subsection{Overall Performance Evaluation}
We mainly compare our algorithm Maple with two recent algorithms, KpLeX (\citeauthor{jiang2021new} \citeyear{jiang2021new})\footnote{\url{https://github.com/huajiang-ynu/kplex}} and kPlexS (\citeauthor{chang2022efficient} \citeyear{chang2022efficient})\footnote{\url{https://lijunchang.github.io/Maximum-kPlex}}.
To the best of our knowledge, KpLeX and kPlexS are state-of-the-art algorithms and dominate earlier algorithms in experiments. Note that, we adapt KpLeX such that it searches for a maximum $k$-plex of size at least $2k-1$.

\subsubsection{Datasets}

We evaluate the algorithms on benchmarks that have been widely used in the literature \citep{gao2018exact,zhou2021improving,jiang2021new,chang2022efficient}.
\begin{itemize}
    \item \textbf{Network-Repo Graphs.} This dataset contains 139 real-world graphs with up to $5.87 \times 10^7$ vertices from the Network Data Repository\footnote{\url{http://lcs.ios.ac.cn/~caisw/Resource/realworld\%20graphs.tar.gz}}, including social networks, biological networks, collaboration networks and so on.
    \item \textbf{10th-DIMACS Graphs.} This dataset contains 84 graphs with up to $5.09 \times 10^7$ vertices\footnote{\url{https://networkrepository.com/dimacs10.php}}, most of them are real-world graphs.
    \item \textbf{2nd-DIMACS Graphs} This dataset contains 80 graphs with up to $4.00 \times 10^3$ vertices\footnote{\url{https://networkrepository.com/dimacs.php}}. Since many graphs of this set are artificial and dense graphs, this set is hard to solve, according to \cite{jiang2021new}.
\end{itemize}

\begin{comment}
To be concise and informative, extremely easy or hard graphs are removed from the presentation and we show the experimental results ono these graphs in the supplement. 
\end{comment}

To be concise and informative, extremely easy or hard graphs are removed from the presentation, and we show the complete experimental results in the supplement. 
Specifically, easy graphs are those that can be solved by kPlexS and Maple for all tested values of $k$ in 5 seconds, and hard graphs are those that cannot be solved in the given time limit of 1800 seconds by kPlexS and Maple for both $k=15$ and $20$.  For several graphs, the optimal solution is smaller than $2k-1$ for both $k=15$ and $20$. These graphs are also removed. The computational results on the remaining 28 Network-Repo graphs, 6 10th-DIMACS graphs, and 14 2nd-DIMACS graphs are presented in Tables \ref{table_overall1} and \ref{table_overall2}. We  use  $T_X$, $T_S$, and $T_M$ to represent the running time of KpLeX, kPlexS and Maple, respectively, OOT to represent \textit{out of time}. 

\subsubsection{Evaluation on Real-World Graphs.}
We first evaluate the algorithms on the real-world graphs, i.e., Network-Repo and 10th-DIMACS graphs.
As shown in the first two blocks of Tables \ref{table_overall1} and \ref{table_overall2}, our algorithm is generally competitive with the reference algorithms for all $k$ values on these real-world graphs.
Specifically, when $k=2,5$, our algorithm is on par with kPlexS, as the performance of kPlexS and Maple contrasts in different scenarios. For example, when $k=5$, Maple achieves an 85x speedup over kPlexS on the soc-flixster graph, while kPlexS achieves a 15x speedup over Maple on the sc-pkustk11 graph.  
But as $k$ becomes larger, the superiority and dominance of our algorithm becomes apparent.
For example, Maple shows a nearly 100x speedup over kPlexS on the sc-pkustk11 graph when $k=15$, and only Maple solves the graphs socfb-Duke14, soc-LiveMocha, soc-youtube, soc-lastfm and consph  in the given time limit when $k=20$.
On the contrary, KpLeX is not as time efficient as kPlexS and Maple in these real-word graphs, and does not scale well for large $k$ values.

\subsubsection{Evaluation on Artificial Dense Graphs.}
We also tested the algorithms on traditional clique graphs, i.e., the 2nd-DIMACS graphs.
As shown in the last block of Tables \ref{table_overall1} and \ref{table_overall2}, the situation is different from real-world graphs. When $k=2,5$, KpLeX performs slightly better than both kPlexS and our algorithm. And when $k=10$, these algorithms compete with each other. However, when $k$ becomes $15$ and $20$, our algorithm still clearly outperforms the others.

In sum, the results indicate that both Maple and kPlexS are scalable to large real-world graphs and large $k$ values, while KpLeX is favorable for solving dense artificial graphs with relatively small $k$ values.

\renewcommand\arraystretch{0.8}

\begin{table}[htbp!]
  \centering
  \caption{Computation results of KpLeX ($T_X$), kPlexS ($T_S$) and Maple ($T_M$) on Network-Repo, 10th-DIMACS and 2nd-DIMACS graphs, with $k=2,5,10,15,20$ and time limit of $1800$ seconds (part I).}
  \resizebox{\linewidth}{!}{
    \centering
    \begin{tabular}{ p{1cm}<{\centering} | p{3.4cm}<{\centering} p{1.8cm}<{\centering} p{2cm}<{\centering} |  
    p{1.4cm}<{\centering} p{1.4cm}<{\centering} p{1.4cm}<{\centering} p{1.4cm}<{\centering}  | p{1.4cm}<{\centering} p{1.4cm}<{\centering} p{1.4cm}<{\centering} p{1.4cm}<{\centering}}
    \toprule
    \multirow{2}{*}{ID} & \multirow{2}{*}{Graph} & \multirow{2}{*}{$|V|$} & \multirow{2}{*}{$|E|$} & \multicolumn{4}{c|}{k=2}                      & \multicolumn{4}{c}{k=5} \\
\cline{5-12}    & & & & $\omega_k(G)$ & $T_{X}$ & $T_{S}$ & $T_{M}$ & $\omega_k(G)$ & $T_{X}$ & $T_{S}$ & $T_{M}$ \\

\hline
G1    & scc\_reality & 6,809 & 4,714,485 & 1236  & \textbf{0.07 } & 0.08  & 0.80  & 1237  & 102.76  & 18.64  & \textbf{13.58 } \\
G2    & tech-WHOIS & 7,476 & 56,943 & 64    & 296.26  & 6.67  & \textbf{0.22 } & 76    & 284.90  & 3.25  & \textbf{0.51 } \\
G3    & socfb-Duke14 & 9,885 & 506,437 & 38    & 248.48  & 54.69  & \textbf{4.61 } & 48    & OOT   & 702.81  & \textbf{9.35 } \\
G4    & soc-epinions & 26,588 & 100,120 & 18    & 0.07  & 0.04 & \textbf{0.04}  & 25    & OOT   & 0.03 & \textbf{0.03}  \\
G5    & socfb-Indiana & 29,732 & 1305,757 & 51    & 284.86  & \textbf{2.23 } & 2.61  & 59    & OOT   & \textbf{2.18 } & 2.31  \\
G6    & ia-email-EU & 32,430 & 54,397 & 15    & 0.03  & 0.02  & \textbf{0.01 } & 20    & 0.05  & 0.01  & \textbf{0.01 } \\
G7    & ia-enron-large & 33,696 & 180,811 & 22    & 0.43  & \textbf{0.09 } & 0.10  & 28    & 10.91  & \textbf{0.08 } & 0.09  \\
G8    & socfb-Texas84 & 36,364 & 1590,651 & 55    & 345.57  & 5.36  & \textbf{2.54 } & 68    & 499.56  & \textbf{2.02 } & 2.13  \\
G9    & sc-nasasrb & 54,870 & 1311,227 & 24    & 9.76  & \textbf{0.57 } & 0.58  & 24    & 149.48  & 3.27  & \textbf{2.35 } \\
G10   & soc-slashdot & 70,068 & 358,647 & 31    & 2.05  & 1.59  & \textbf{0.30 } & 40    & OOT   & 1.21  & \textbf{0.21 } \\
G11   & sc-pkustk11 & 87,804 & 2,565,054 & 36    & \textbf{0.70 } & 1.36  & 1.28  & 36    & \textbf{2.70 } & 10.79  & 161.35  \\
G12   & ia-wiki-Talk & 92,117 & 360,767 & 18    & \textbf{1.51 } & 3.84  & 2.52  & 25    & OOT   & 4.84  & \textbf{2.50 } \\
G13   & soc-LiveMocha & 104,103 & 2,193,083 & 19    & 18.89  & 64.11  & \textbf{15.67 } & 28    & OOT   & OOT   & \textbf{128.50 } \\
G14   & soc-gowalla & 196,591 & 950,327 & 30    & 2.10  & \textbf{0.32 } & 0.35  & 32    & OOT   & \textbf{0.26 } & 0.29  \\
G15   & sc-pwtk & 217,891 & 5,653,221 & 24    & OOT   & 2.31  & \textbf{2.25 } & 26    & OOT   & 4.38  & \textbf{4.32 } \\
G16   & soc-twitter-follows & 404,719 & 713,319 & 8     & 1.11  & \textbf{0.13 } & 0.16  & 13    & OOT   & 0.12  & \textbf{0.11 } \\
G17   & sc-msdoor & 404,785 & 9,378,650 & 21    & OOT   & 13.32  & \textbf{13.08 } & 23    & OOT   & \textbf{456.11 } & 630.03  \\
G18   & soc-youtube & 495,957 & 1,936,748 & 20    & 1.39  & \textbf{0.75 } & 0.79  & 26    & 57.92  & 0.79  & \textbf{0.74 } \\
G19   & soc-FourSquare & 639,014 & 3,214,986 & 35    & 969.42  & \textbf{16.97 } & 23.64  & 44    & OOT   & \textbf{3.88 } & 9.28  \\
G20   & soc-digg & 770,799 & 5,907,132 & 57    & OOT   & OOT   & \textbf{572.39 } & 72    & OOT   & OOT   & OOT \\
G21   & sc-ldoor & 909,537 & 20,770,807 & 21    & OOT   & 22.80  & \textbf{21.23 } & 23    & OOT   & \textbf{820.11 } & 1056.57  \\
G22   & soc-youtube-snap & 1,134,890 & 2,987,624 & 20    & 1.96  & \textbf{1.08 } & 1.25  & 26    & 147.49  & \textbf{0.96 } & 1.04  \\
G23   & soc-lastfm & 1,191,805 & 4,519,330 & 18    & \textbf{8.81 } & 39.58  & 14.75  & 27    & OOT   & 143.43  & \textbf{65.42 } \\
G24   & soc-pokec & 1,632,803 & 22,301,964 & 31    & \textbf{16.74 } & 30.24  & 31.89  & 34    & OOT   & \textbf{25.59 } & 28.20  \\
G25   & soc-flixster & 2,523,386 & 7,918,801 & 38    & 36.31  & 39.09  & \textbf{2.40 } & 49    & OOT   & 190.51  & \textbf{2.21 } \\
G26   & socfb-B-anon & 2,937,612 & 20,959,854 & 27    & 146.31  & \textbf{38.22 } & 38.46  & 35    & OOT   & \textbf{42.63 } & 46.78  \\
G27   & soc-orkut & 2,997,166 & 106,349,209 & 52    & OOT   & OOT   & \textbf{398.48 } & 68    & OOT   & OOT   & OOT \\
G28   & socfb-A-anon & 3,097,165 & 23,667,394 & 28    & \textbf{30.10 } & 32.32  & 35.54  & 37    & 602.96  & 33.63  & \textbf{32.50 } \\
\hline
G29   & consph & 79,679  & 2,963,573  & 24    & 403.20  & \textbf{1.75 } & 2.05  & 26    & 560.42  & 56.14  & \textbf{26.75 } \\
G30   & connectus & 394,707  & 1,127,491  & 12    & 1.95  & \textbf{0.24 } & 0.36  & 19    & 0.76  & \textbf{0.25 } & 0.26  \\
G31   & rgg\_n\_2\_21\_s0 & 2,097,148  & 14,487,995  & 19    & 1.52  & 1.24  & \textbf{0.31 } & 22    & 1.77  & \textbf{0.42 } & 0.43  \\
G32   & rgg\_n\_2\_22\_s0 & 4,194,301  & 30,359,198  & 20    & 3.98  & 2.60  & \textbf{0.73 } & 23    & 4.15  & \textbf{0.90 } & 1.00  \\
G33   & rgg\_n\_2\_23\_s0 & 8,388,607  & 63,501,393  & 22    & 8.32  & 5.12  & \textbf{1.65 } & 24    & 9.19  & 2.52  & \textbf{2.21 } \\
G34   & rgg\_n\_2\_24\_s0 & 16,777,215  & 132,557,200  & 22    & 18.30  & 10.89  & \textbf{4.64 } & 25    & 20.73  & \textbf{4.14 } & 4.30  \\
\hline
G35   & hamming6-4 & 64    & 704   & 6     & 0.01  & 0.01  & \textbf{0.01 } & 12    & 0.04  & \textbf{0.01 } & 0.03  \\
G36   & hamming6-2 & 64    & 1,824  & 32    & 18.21  & 17.84  & \textbf{0.71 } & 48    & 1567.71  & 1357.46  & \textbf{86.86 } \\
G37   & johnson8-4-4 & 70    & 1,855  & 14    & \textbf{2.58 } & 17.05  & 3.38  & 28    & \textbf{407.74 } & OOT   & OOT \\
G38   & C125.9 & 125   & 6,963  & OOT   & OOT   & OOT   & OOT   & OOT   & OOT   & OOT   & OOT \\
G39   & san200\_0.7\_2 & 200   & 13,930  & OOT   & OOT   & OOT   & OOT   & OOT   & OOT   & OOT   & OOT \\
G40   & san200\_0.7\_1 & 200   & 13,930  & OOT   & OOT   & OOT   & OOT   & OOT   & \textbf{1309.52 } & OOT   & OOT \\
G41   & hamming8-2 & 256   & 31,616  & OOT   & OOT   & OOT   & OOT   & OOT   & OOT   & OOT   & OOT \\
G42   & MANN\_a27 & 378   & 70,551  & 236   & OOT   & OOT   & \textbf{629.38 } & 351   & \textbf{0.21 } & 0.94  & 0.86  \\
G43   & san400\_0.7\_2 & 400   & 55,860  & OOT   & OOT   & OOT   & OOT   & OOT   & OOT   & OOT   & OOT \\
G44   & san400\_0.7\_3 & 400   & 55,860  & OOT   & OOT   & OOT   & OOT   & OOT   & OOT   & OOT   & OOT \\
G45   & san400\_0.7\_1 & 400   & 55,860  & OOT   & OOT   & OOT   & OOT   & OOT   & OOT   & OOT   & OOT \\
G46   & c-fat500-2 & 500   & 9,139  & 26    & \textbf{0.00 } & 0.01  & 0.01  & 26    & 0.03  & \textbf{0.00 } & 0.01  \\
G47   & hamming10-2 & 1,024  & 518,656  & OOT   & OOT   & OOT   & OOT   & OOT   & OOT   & OOT   & OOT \\
G48   & MANN\_a45 & 1,035  & 533,115  & OOT   & OOT   & OOT   & OOT   & 990   & \textbf{1.58 } & 21.70  & 22.06  \\
\bottomrule

    \end{tabular}%
}
  \label{table_overall1}%
\end{table}%

\begin{table}[htbp!]
  \centering
  \caption{Computation results of KpLeX ($T_X$), kPlexS ($T_S$) and Maple ($T_M$) on Network-Repo, 10th-DIMACS and 2nd-DIMACS graphs, with $k=2,5,10,15,20$ and time limit  of $1800$ seconds (part II).
  }
  \resizebox{\linewidth}{!}{
    \centering
    \begin{tabular}{
    p{1cm}<{\centering} |
    p{1.5cm}<{\centering} p{1.5cm}<{\centering} p{1.5cm}<{\centering} p{1.5cm}<{\centering}|  
    p{1.5cm}<{\centering} p{1.5cm}<{\centering} p{1.5cm}<{\centering} p{1.5cm}<{\centering}| p{1.5cm}<{\centering} p{1.5cm}<{\centering} p{1.5cm}<{\centering} p{1.5cm}<{\centering}}
    \toprule
\multirow{2}{*}{ID} & \multicolumn{4}{c|}{k=10} & \multicolumn{4}{c|}{k=15} &
\multicolumn{4}{c}{k=20} \\
\cline{2-13} & $\omega_k(G)$ & $T_{X}$ & $T_{S}$ & $T_{M}$ & $\omega_k(G)$ & $T_{X}$ & $T_{S}$ & $T_{M}$ & $\omega_k(G)$ & $T_{X}$ & $T_{S}$ & $T_{M}$ \\

\hline
G1    & 1239  & \textbf{5.16 } & 5.83  & 7.97  & 1244  & OOT   & \textbf{58.27 } & 70.30  & 1251  & OOT   & \textbf{49.87 } & 50.09  \\
G2    & 87    & 552.64  & 0.17  & \textbf{0.11 } & 96    & 29.50  & 0.40  & \textbf{0.30 } & 104   & 3.08  & 0.11  & \textbf{0.08 } \\
G3    & 60    & OOT   & 298.91  & \textbf{5.31 } & 70    & OOT   & OOT   & \textbf{32.81 } & 81    & OOT   & OOT   & \textbf{43.75 } \\
G4    & 33    & 0.03  & 0.02  & \textbf{0.02 } & 40    & 5.08  & 0.50  & \textbf{0.23 } & 46    & OOT   & 57.31  & \textbf{10.47 } \\
G5    & 70    & 723.78  & \textbf{1.56 } & 1.68  & 75    & OOT   & \textbf{1.42 } & 1.60  & 83    & 446.69  & 14.32  & \textbf{5.30 } \\
G6    & 26    & 0.31  & 0.04  & \textbf{0.03 } & 33    & 33.93  & 0.18  & \textbf{0.18 } & 39    & OOT   & \textbf{2.26 } & 5.69  \\
G7    & 38    & OOT   & \textbf{0.08 } & 0.09  & 45    & OOT   & \textbf{0.12 } & 0.14  & 51    & OOT   & 8.21  & \textbf{2.12 } \\
G8    & 79    & \textbf{0.58 } & 1.24  & 1.38  & 87    & 10.79  & \textbf{1.06 } & 1.14  & 94    & 2.99  & \textbf{0.77 } & 0.79  \\
G9    & 31    & OOT   & 2.22  & \textbf{1.94 } & 36    & OOT   & 148.89  & \textbf{13.54 } & 42    & OOT   & 459.33  & \textbf{12.31 } \\
G10   & 51    & OOT   & 0.13  & \textbf{0.07 } & 59    & OOT   & 9.47  & \textbf{1.25 } & 68    & 110.08  & \textbf{0.11 } & 0.27  \\
G11   & 48    & 431.69  & 397.89  & \textbf{96.99 } & 48    & OOT   & 136.90  & \textbf{1.41 } & 56    & OOT   & 407.64  & \textbf{14.35 } \\
G12   & 35    & OOT   & 6.86  & \textbf{4.62 } & 44    & OOT   & 126.63  & \textbf{16.49 } & OOT   & OOT   & OOT   & OOT \\
G13   & 41    & OOT   & 628.76  & \textbf{49.18 } & 52    & OOT   & 3.76  & \textbf{3.65 } & 60    & OOT   & OOT   & \textbf{595.15 } \\
G14   & 42    & OOT   & 0.26  & \textbf{0.23 } & 49    & OOT   & 0.39 & \textbf{0.39}  & 56    & OOT   & 64.12  & \textbf{7.98 } \\
G15   & 33    & OOT   & \textbf{10.85 } & 11.09  & 38    & OOT   & 116.60  & \textbf{20.08 } & 46    & OOT   & 122.78  & \textbf{6.31 } \\
G16   & 21    & OOT   & \textbf{0.11 } & 0.18  & 30    & OOT   & 5.60  & \textbf{0.12 } & 38    & OOT   & 24.93  & \textbf{20.11 } \\
G17   & 35    & OOT   & \textbf{5.14 } & 6.02  & 42    & OOT   & \textbf{3.83 } & 3.88  & 45    & OOT   & 887.96  & \textbf{17.36 } \\
G18   & 35    & OOT   & 0.64 & \textbf{0.64}  & 43    & OOT   & 16.30  & \textbf{5.16 } & 50    & OOT   & OOT   & \textbf{651.92 } \\
G19   & 53    & OOT   & \textbf{4.33 } & 5.80  & 59    & OOT   & 7.70  & \textbf{7.23 } & 65    & OOT   & 8.98  & \textbf{7.75 } \\
G20   & 87    & OOT   & \textbf{928.29 } & 1045.20  & 100   & OOT   & \textbf{14.85 } & 19.62  & 109   & OOT   & 15.81  & \textbf{14.57 } \\
G21   & 35    & OOT   & 11.26  & \textbf{11.14 } & 42    & OOT   & \textbf{7.91 } & 9.17  & 45    & OOT   & 14.11  & \textbf{12.78 } \\
G22   & 35    & OOT   & 1.09  & \textbf{0.96 } & 43    & OOT   & 33.11  & \textbf{7.62 } & 51    & OOT   & OOT   & OOT \\
G23   & 38    & OOT   & 5.38  & \textbf{2.95 } & 47    & OOT   & 84.67  & \textbf{12.42 } & 56    & OOT   & OOT   & \textbf{1417.15 } \\
G24   & 45    & 1451.55  & \textbf{20.40 } & 22.70  & 49    & OOT   & \textbf{16.01 } & 23.59  & 55    & OOT   & 14.99  & \textbf{14.22 } \\
G25   & 62    & OOT   & 20.99  & \textbf{1.63 } & 72    & 259.17  & 1.11  & \textbf{0.45 } & 81    & 1.41  & 0.54  & \textbf{0.30 } \\
G26   & 47    & OOT   & \textbf{28.19 } & 31.88  & 57    & OOT   & \textbf{24.92 } & 30.86  & 64    & OOT   & 27.28  & \textbf{21.99 } \\
G27   & 89    & OOT   & OOT   & \textbf{317.00 } & 101   & OOT   & \textbf{213.18 } & 245.11  & 111   & OOT   & 181.54  & \textbf{179.85 } \\
G28   & 47    & OOT   & \textbf{24.49 } & 27.94  & 54    & OOT   & \textbf{21.67 } & 26.94  & 61    & OOT   & 25.05  & \textbf{18.54 } \\
\hline
G29   & 33    & OOT   & 19.27  & \textbf{9.10 } & 42    & OOT   & 437.79  & \textbf{20.25 } & 45    & OOT   & OOT   & \textbf{327.47 } \\
G30   & 26    & OOT   & \textbf{0.82 } & 1.06  & 34    & OOT   & 77.09  & \textbf{69.05 } & OOT   & OOT   & OOT   & OOT \\
G31   & 25    & OOT   & \textbf{2.55 } & 2.56  & 30    & OOT   & 6.35  & \textbf{0.44 } & 38    & 1.71  & 1.39  & \textbf{0.26 } \\
G32   & 27    & OOT   & 3.95  & \textbf{3.67 } & 31    & OOT   & 13.99  & \textbf{1.03 } & 38    & OOT   & 14.64  & \textbf{0.59 } \\
G33   & 28    & 10.07  & \textbf{5.69 } & 6.22  & 33    & OOT   & 30.35  & \textbf{2.33 } & 38    & OOT   & 31.87  & \textbf{1.32 } \\
G34   & 29    & 27.52  & 13.78  & \textbf{11.05 } & 33    & OOT   & 28.13  & \textbf{5.40 } & 38    & OOT   & 67.92  & \textbf{2.93 } \\
\hline
G35   & 20    & 0.54  & 0.01  & \textbf{0.01 } & 30    & \textbf{0.00 } & 0.00  & 0.01  & 38    & 0.22  & OOT   & \textbf{0.01 } \\
G36   & 64    & \textbf{0.00 } & 0.00  & 0.00  & 64    & 0.38  & 0.00 & \textbf{0.00}  & 64    & \textbf{0.00 } & 0.00  & 0.00  \\
G37   & OOT   & OOT   & OOT   & OOT   & 60    & OOT   & 676.19  & \textbf{90.72 } & 70    & \textbf{0.00 } & 0.00  & 0.00  \\
G38   & OOT   & OOT   & OOT   & OOT   & 112   & OOT   & 158.44  & \textbf{76.98 } & 122   & \textbf{0.00 } & 0.00  & 0.00  \\
G39   & OOT   & \textbf{0.62 } & 2.54  & OOT   & 134   & OOT   & 0.00 & \textbf{0.00}  & 134   & \textbf{0.00 } & 0.00  & 0.00  \\
G40   & 105   & \textbf{0.02 } & 0.03  & 0.03  & 105   & OOT   & 0.03  & \textbf{0.03 } & 105   & OOT   & OOT   & \textbf{98.28 } \\
G41   & 256   & \textbf{0.00 } & 0.01  & 0.01  & 256   & OOT   & 0.01 & \textbf{0.01}  & 256   & \textbf{0.00 } & 0.01  & 0.01  \\
G42   & 351   & \textbf{2.62 } & OOT   & 280.40  & 378   & OOT   & 0.01 & \textbf{0.01}  & 378   & \textbf{0.00 } & 0.01  & 0.01  \\
G43   & OOT   & OOT   & OOT   & OOT   & 205   & OOT   & \textbf{0.19 } & 0.22  & 205   & \textbf{0.14 } & 0.18  & 0.17  \\
G44   & OOT   & OOT   & OOT   & OOT   & OOT   & OOT   & OOT   & OOT   & 216   & \textbf{0.11 } & 0.18  & 0.19  \\
G45   & 200   & 0.16  & 0.17  & \textbf{0.15 } & 200   & OOT   & \textbf{0.19 } & 0.20  & 200   & 0.20  & 0.19  & \textbf{0.18 } \\
G46   & 31    & 0.03  & 0.01  & \textbf{0.01 } & 39    & 59.86  & 0.01  & \textbf{0.01 } & 39    & OOT   & 13.65  & \textbf{0.01 } \\
G47   & OOT   & OOT   & OOT   & OOT   & 1024  & OOT   & \textbf{0.06 } & 0.07  & 1024  & \textbf{0.00 } & 0.06  & 0.07  \\
G48   & 990   & \textbf{4.73 } & OOT   & 90.35  & 990   & OOT   & OOT   & \textbf{475.88 } & OOT   & OOT   & OOT   & OOT \\
\bottomrule

    \end{tabular}%
}
  \label{table_overall2}%
\end{table}%

\subsection{Analysis on Key Components}
\begin{table}[htbp!]
  \centering
  \caption{Computation time of Maple ($T_M$), Maple$\mathrm{\mathbf{_{com}}}$ ($T_C$) and Maple$\mathrm{_{\mathbf{hyb}}}$ ($T_H$) on instances with high degeneracy but low degeneracy gap, i.e., $d(G) \ge 50$ and $g_k(G) \le 10$.}
  \resizebox{\linewidth}{!}{
    \begin{tabular}{
    p{3.6cm}<{\centering} p{1.6cm}<{\centering} p{1.9cm}<{\centering} p{1.3cm}<{\centering} p{1.3cm}<{\centering} p{1.3cm}<{\centering} p{1.3cm}<{\centering} p{1.3cm}<{\centering} p{1.3cm}<{\centering} p{1.3cm}<{\centering} p{1.3cm}<{\centering} p{1.3cm}<{\centering}
    }
    \toprule
    Graph & $|V|$ & $|E|$ & $d(G)$ & $cd(G)$ & $k$ & $\omega_k(G)$ & $g_k(G)$ & $cg_k(G)$ & $T_M$ & $T_C$ & $T_H$ \\
    \hline
    scc\_infect-hyper & 113 & 6222 & 105 & 104 & 2 & 106 & 1 & 2 & 0 & 0 & 0 \\ 
    scc\_infect-hyper & 113 & 6222 & 105 & 104 & 5 & 107 & 3 & 7 & 0 & 0 & 0 \\ 
    scc\_enron-only & 146 & 9828 & 119 & 118 & 2 & 121 & 0 & 1 & 0 & 0 & 0 \\ 
    scc\_enron-only & 146 & 9828 & 119 & 118 & 5 & 123 & 1 & 5 & 0.02 & 0.02 & 0.02 \\ 
    scc\_fb-forum & 488 & 71011 & 272 & 264 & 2 & 266 & 8 & 2 & 0.23 & 0.2 & 0.31 \\ 
    scc\_fb-forum & 488 & 71011 & 272 & 264 & 5 & 272 & 5 & 2 & 0.55 & 0.28 & 0.63 \\ 
    scc\_fb-forum & 488 & 71011 & 272 & 264 & 10 & 281 & 1 & 3 & 0.02 & 0.02 & 0.02 \\ 
    scc\_fb-messages & 1303 & 531893 & 706 & 705 & 2 & 708 & 0 & 1 & 0.11 & 0.1 & 0.09 \\ 
    scc\_fb-messages & 1303 & 531893 & 706 & 705 & 5 & 709 & 2 & 6 & 0.11 & 0.1 & 0.1 \\ 
    scc\_twitter-copen & 2623 & 473614 & 582 & 579 & 2 & 581 & 3 & 2 & 0.92 & 2.87 & 2.11 \\ 
    scc\_reality & 6809 & 4714485 & 1235 & 1234 & 2 & 1236 & 1 & 2 & 0.8 & 0.77 & 0.72 \\ 
    scc\_reality & 6809 & 4714485 & 1235 & 1234 & 5 & 1237 & 3 & 7 & 13.58 & 12.39 & 13.39 \\ 
    tech-WHOIS & 7476 & 56943 & 88 & 69 & 15 & 96 & 7 & 3 & 0.3 & 0.17 & 0.3 \\ 
    tech-WHOIS & 7476 & 56943 & 88 & 69 & 20 & 104 & 4 & 5 & 0.08 & 0.36 & 0.08 \\ 
    scc\_infect-dublin & 10972 & 175573 & 83 & 82 & 2 & 84 & 1 & 2 & 0 & 0 & 0 \\ 
    ca-HepPh & 11204 & 117619 & 238 & 237 & 2 & 239 & 1 & 2 & 0.01 & 0.01 & 0.01 \\ 
    socfb-UCSB37 & 14917 & 482215 & 65 & 58 & 5 & 68 & 2 & 0 & 0.09 & 0.08 & 0.08 \\ 
    socfb-UCSB37 & 14917 & 482215 & 65 & 58 & 10 & 75 & 0 & 3 & 0.02 & 0.02 & 0.02 \\ 
    ca-AstroPh & 17903 & 196972 & 56 & 55 & 2 & 57 & 1 & 2 & 0.01 & 0.01 & 0.01 \\ 
    socfb-UCLA & 20453 & 747604 & 65 & 52 & 5 & 62 & 8 & 0 & 0.83 & 0.79 & 0.79 \\ 
    socfb-UF & 35111 & 1465654 & 83 & 65 & 20 & 99 & 4 & 6 & 0.67 & 0.73 & 0.78 \\ 
    soc-brightkite & 56739 & 212945 & 52 & 41 & 5 & 51 & 6 & 0 & 0.02 & 0.01 & 0.02 \\ 
    soc-brightkite & 56739 & 212945 & 52 & 41 & 10 & 58 & 4 & 3 & 0.02 & 0.03 & 0.02 \\ 
    soc-brightkite & 56739 & 212945 & 52 & 41 & 15 & 65 & 2 & 6 & 0.01 & 0.01 & 0.01 \\ 
    socfb-OR & 63392 & 816886 & 52 & 34 & 10 & 53 & 9 & 1 & 0.29 & 0.3 & 0.35 \\ 
    soc-slashdot & 70068 & 358647 & 53 & 33 & 20 & 68 & 5 & 5 & 0.27 & 1.44 & 0.08 \\ 
    web-sk-2005 & 121422 & 334419 & 81 & 80 & 2 & 82 & 1 & 2 & 0.01 & 0.02 & 0.02 \\ 
    web-sk-2005 & 121422 & 334419 & 81 & 80 & 5 & 83 & 3 & 7 & 0.02 & 0.02 & 0.02 \\ 
    web-uk-2005 & 129632 & 11744049 & 499 & 498 & 2 & 500 & 1 & 2 & 0.67 & 1.77 & 1.88 \\ 
    web-arabic-2005 & 163598 & 1747269 & 101 & 100 & 2 & 102 & 1 & 2 & 0.01 & 0.01 & 0.01 \\ 
    ca-dblp-2010 & 226413 & 716460 & 74 & 73 & 2 & 75 & 1 & 2 & 0.02 & 0.02 & 0.02 \\ 
    ca-citeseer & 227320 & 814134 & 86 & 85 & 2 & 87 & 1 & 2 & 0.03 & 0.03 & 0.03 \\ 
    ca-dblp-2012 & 317080 & 1049866 & 113 & 112 & 2 & 114 & 1 & 2 & 0.02 & 0.02 & 0.02 \\ 
    web-it-2004 & 509338 & 7178413 & 431 & 430 & 2 & 432 & 1 & 2 & 0.06 & 0.06 & 0.06 \\ 
    ca-coauthors-dblp & 540486 & 15245729 & 336 & 335 & 2 & 337 & 1 & 2 & 0.13 & 0.12 & 0.12 \\ 
    ca-hollywood-2009 & 1069126 & 56306653 & 2208 & 2207 & 2 & 2209 & 1 & 2 & 1.49 & 0.99 & 0.94 \\ 
    soc-livejournal & 4033137 & 27933062 & 213 & 212 & 2 & 214 & 1 & 2 & 1.75 & 1.85 & 1.85 \\ \hline
    coAuthorsCiteseer & 227320 & 814134 & 86 & 85 & 2 & 87 & 1 & 2 & 0.03 & 0.03 & 0.03 \\ 
    cnr-2000 & 325557 & 2738969 & 83 & 82 & 2 & 85 & 0 & 1 & 0.09 & 0.08 & 0.07 \\ 
    cnr-2000 & 325557 & 2738969 & 83 & 82 & 5 & 86 & 2 & 6 & 0.09 & 0.08 & 0.09 \\ 
    co-papers-citeseer & 434102 & 16036720 & 844 & 843 & 2 & 845 & 1 & 2 & 0.11 & 0.11 & 0.1 \\ 
    co-papers-dblp & 540486 & 15245729 & 336 & 335 & 2 & 337 & 1 & 2 & 0.13 & 0.11 & 0.12 \\  \hline
    hamming6-2 & 64 & 1824 & 57 & 50 & 10 & 64 & 3 & 6 & 0 & 0 & 0 \\ 
    johnson8-4-4 & 70 & 1855 & 53 & 36 & 20 & 70 & 3 & 6 & 0 & 0 & 0 \\ 
    C125.9 & 125 & 6963 & 102 & 84 & 15 & 112 & 5 & 2 & 76.98 & 5.31 & 83.42 \\ 
    C125.9 & 125 & 6963 & 102 & 84 & 20 & 122 & 0 & 2 & 0 & 0 & 0 \\ 
    san200\_0.7\_2 & 200 & 13930 & 122 & 110 & 15 & 134 & 3 & 6 & 0 & 0 & 0 \\ 
    hamming8-2 & 256 & 31616 & 247 & 238 & 10 & 256 & 1 & 2 & 0.01 & 0.01 & 0.01 \\ 
    MANN\_a27 & 378 & 70551 & 364 & 350 & 15 & 378 & 1 & 2 & 0.01 & 0.01 & 0.01 \\ 
    \bottomrule
    \end{tabular}
    \label{table_low_high}
}
\end{table}

Our algorithm design and implementation include several key components such as the degeneracy gap, the graph reduction and the bounding techniques. 
We perform a breakup analysis to investigate the influence of these components. For comparison, different versions of our main algorithm Maple are tested in this section.
\begin{itemize}
    \item Maple - Our main algorithm equipped with KPLEX in Alg. \ref{Alg_framework}, where the search space is decomposed with respect to vertices.
    \item Maple$\mathrm{_{com}}$ - The variant equipped with KPLEX$\mathrm{_{com}}$ in Alg. \ref{Alg_framework2} instead of KPLEX, where the search space is decomposed with respect to edges.
    \item Maple$\mathrm{_{hyb}}$ - The variant as an exploratory hybrid of Maple and Maple$\mathrm{_{com}}$, where the search space is decomposed with respect to vertices or edges dynamically. That is, the optimal vertex $v$ or edge $e$ to anchor the search subspace is selected alternately under the following rule: choose vertex $v$ if $N(v)+k \le N(e)+2k$, otherwise choose edge $e$. The underlying principle of our rule is to minimize the current search subspace, and the details of Maple$\mathrm{_{hyb}}$ is given in the supplement.
    \item Maple$\mathrm{_{NGR}}$ - The variant with the second- and higher-order graph reduction disabled.
    % in Alg. \ref{alg_reduce} 
    \item Maple$\mathrm{_{NBD}}$ - The variant with the bound for the minimum $d$-bdd size disabled.
     % in Alg. \ref{alg_bound}
\end{itemize}

\subsubsection{Analysis on Degeneracy Gap.}
We investigate the degeneracy gap $g_k(G)=d(G)+k-\omega_k(G)$ and the community-degeneracy gap $cg_k(G)=cd(G)+2k-\omega_k(G)$.
In Table \ref{table_low_high}, we present all 49 solved instances with a large degeneracy but a low degeneracy gap, that is, $d(G) \ge 50$ and $g_k(G) \le 10$.

These instances cover all the $k$ values of $2,5,10,15$ and $20$.
For many graphs, the degeneracy can be as large as hundreds. In expectation, solving these graphs is considered extremely difficult because they are parameterized by the degeneracy values for any $k$ values, but in reality most of them can be done in 1 second; even the hardest one takes only about one minute. 
For example, the scc\_reality graph has a degeneracy of $1235$. But when $k=2$, the degeneracy gap of the graph is $1$, which possibly explains why it can be solved in just 1 second.

\begin{figure}[htbp!]
\centering
\includegraphics[width=\columnwidth]{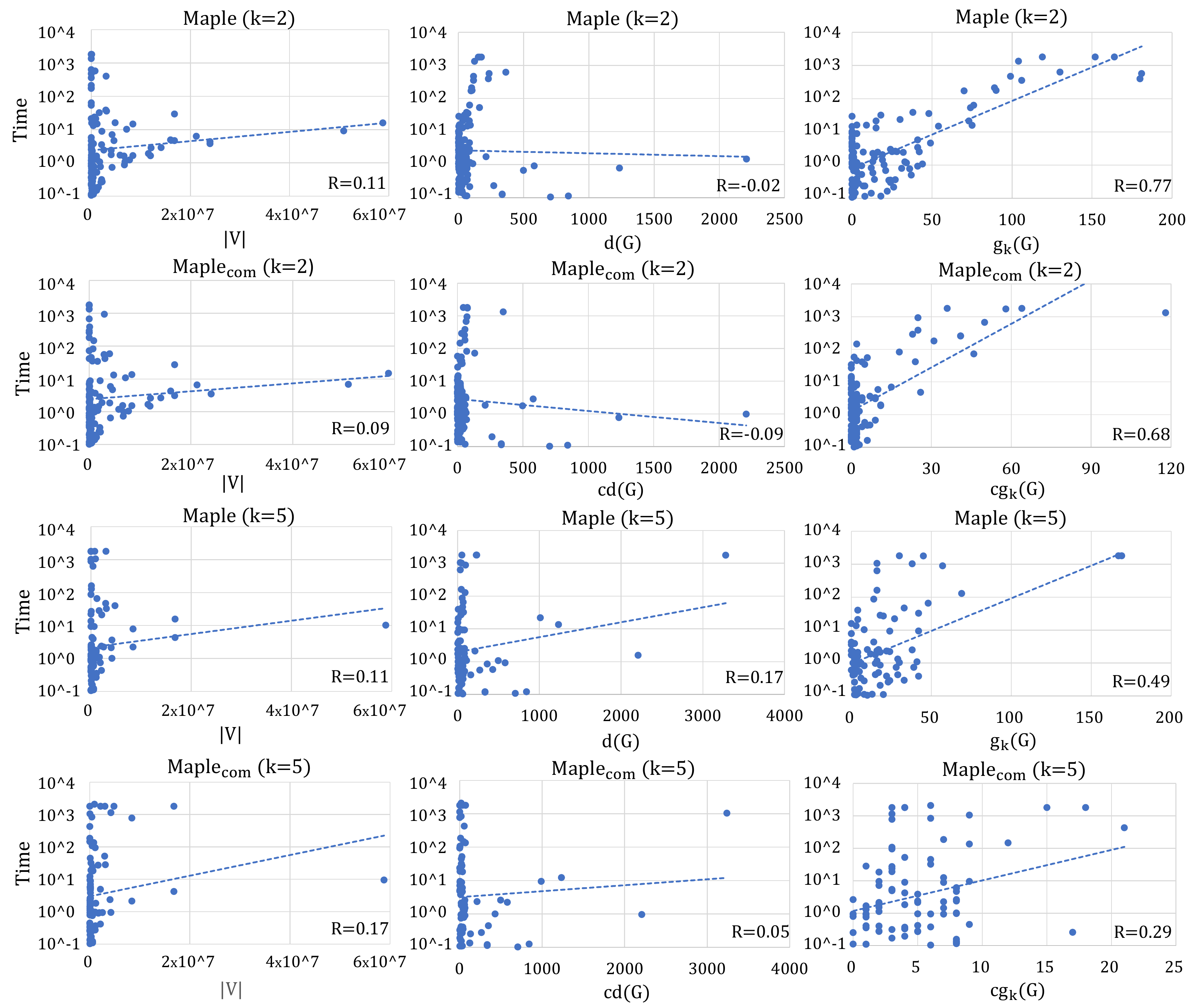}
% \vspace{-4mm}
\caption{Scatter plots of the running time of Maple and Maple$\mathrm{\mathbf{_{com}}}$ against the number of vertices $|V|$, the degeneracy values $d(G)$ and $cd(G)$, and the gap values $g_k(G)$ and $cg_k(G)$, for $k=2,5$.\label{fig-correl}}
\end{figure}

We further investigate the correlation between the graph parameters and our algorithms. 
In Figure \ref{fig-correl}, for Maple and Maple$\mathrm{_{com}}$, we present scatter plots for all the solved instances.
We show the logarithmic running time against the vertex number, the degeneracy parameters, and the gap parameters in each of these plots.
Pearson correlation coefficients are also calculated.
We observed that there is almost no linear dependency between the vertex number and the running time.
Also, no linear dependency exists between the (community-) degeneracy and the running time.
On the flip side, there is generally a positive linear dependency between the (community-) degeneracy gap and the running time.
We also observe that the correlation coefficient between running time and the degeneracy gap tends to decrease as $k$ increases.
Specifically, for $k=2$, the coefficient between time and $g_k(G)$ is 0.77. For $k=5$, the coefficient is 0.49. Meanwhile, the coefficient between time and $cg_k(G)$ changes from 0.68 to 0.29 as $k$ changes from 2 to 5.
This could possibly be explained by our complexity results. That is, the polynomial factor in complexity becomes more influential as $k$ increases. 
As a matter of fact, a perfect correlation is difficult to attain, as the running time is affected by many factors, including graph reduction, bound estimation, and even machine configuration.

In Table \ref{tab-demo}, we also present information on six representative graphs from the aforementioned datasets. 
We notice that, the degeneracy gap is often much smaller than the degeneracy.
For example, $d(G)$ can be as large as 213 in soc-livejournal, while the $g_k(G)$ is only 1 when $k=2$. 
Indeed, our statistic result also reveals that $g_k(G)$ is often bounded by $O(\log{(|V|)})$ in most real-world graphs.
On the flip side, the degeneracy gaps on artificial, dense graphs are large. When $k=2,5,10$, $g_k(G)$ of C125.9 is about $62$, $44$ and $21$, respectively. 
Consequently, our algorithm cannot solve these instances in a reasonable time. 
When $k=15$ and $20$, $g_k(G)$ of C125.9 is $5$ and $0$, respectively, and our algorithm can easily solve it in $66$ and $0$ seconds, respectively.

We also observed that the community-degeneracy $cd(G)$ is constantly smaller than the degeneracy $d(G)$ for all these instances.
However, this does not hold for the community-degeneracy gap and the degeneracy gap.
For small $k$ values like $2,5$, the community-degeneracy gap $cg_k(G)$ is usually smaller than the degeneracy gap $g_k(G)$. But as $k$ increases to $15$ or $20$, the community-degeneracy gap exceeds the degeneracy gap.

If we can compare the performance of our algorithms, Maple with the degeneracy ordering, Maple$\mathrm{_{com}}$ with the community-degeneracy ordering, and Maple$\mathrm{_{hyb}}$ with a hybrid mechanism. The results may vary depending on the scenario.
In general, Maple outperforms Maple$\mathrm{_{hyb}}$ and shows a more robust performance than Maple$\mathrm{_{com}}$. 

\begin{table}[htbp!]
  \centering
  \caption{Properties and running behavior on six representative graphs. 
  $T_M$, $T_C$, $T_H$, $T_{\overline{R}}$ and $T_{\overline{B}}$ stand for the running time of Maple, Maple$\mathrm{\mathbf{_{com}}}$, Maple$\mathrm{\mathbf{_{hyb}}}$, Maple$\mathrm{\mathbf{_{NGR}}}$ and Maple$\mathrm{\mathbf{_{NBD}}}$ respectively. $\gamma$ is the average branching factor of Maple.\label{tab-demo}}
  \resizebox{\linewidth}{!}{
    \begin{tabular}{p{2.4cm}<{\centering} p{1cm}<{\centering} p{1.1cm}<{\centering} p{1.1cm}<{\centering} p{1.1cm}<{\centering} p{1.9cm}<{\centering} p{1.9cm}<{\centering} p{1.9cm}<{\centering} p{1.9cm}<{\centering} p{2cm}<{\centering} p{1cm}<{\centering}
    }
    \toprule
    Graph & $k$ & $\omega_k(G)$ & $g_k(G)$ & $cg_k(G)$ & $T_M$ & $T_C$ & $T_H$  & $T_{\overline{R}}$ & $T_{\overline{B}}$ & $\gamma$ \\
    \hline
    \multirow{5}{*}{\tabincell{c}{ sc-pwtk \\ $|V|=217891 $ \\ $|E|=5653221$ \\ $d(G)=35$ \\ $cd(G)=22$}} 
    & 2     & 24    & 13    & 2     & 2.18  & 8.78  & 3.64  & 3.88  & 2.24  & 2.00  \\
    & 5     & 26    & 14    & 6     & 3.35  & 31.72  & 5.28  & 3.82  & 3.37  & 1.99  \\
    & 10    & 33    & 12    & 9     & 10.43  & 144.63  & 12.05  & 10.80  & 10.57  & 2.00  \\
    & 15    & 38    & 12    & 14    & 18.59  & 116.57  & 25.66  & OOT   & 16.25  & 1.64  \\
    & 20    & 46    & 9     & 16    & 6.22  & 61.78  & 84.10  & 18.20  & 6.23  & 1.97  \\
    \hline
    \multirow{5}{*}{\tabincell{c}{ soc-lastfm \\ $|V|=1191805 $ \\ $|E|=4519330$ \\ $d(G)=70$ \\ $cd(G)=21$}} 
    & 2     & 18    & 54    & 7     & 13.66  & 3.32  & 2.97  & 16.94  & 14.20  & 1.30  \\
    & 5     & 27    & 48    & 4     & 59.99  & 1.95  & 2.94  & 130.05  & 155.09  & 1.49  \\
    & 10    & 38    & 42    & 3     & 2.25  & 1.83  & 1.58  & OOT   & 2.26  & 1.43  \\
    & 15    & 47    & 38    & 4     & 11.76  & 17.69  & 13.62  & OOT   & 11.28  & 1.84  \\
    & 20    & 56    & 34    & 5     & 1412.91  & 174.23  & 1476.13  & OOT   & 1269.92  & 1.80  \\
    \hline
    \multirow{5}{*}{\tabincell{c}{ soc-livejournal \\ $|V|=4033137 $ \\ $|E|=2193083$ \\ $d(G)=213$ \\ $cd(G)=212$}} 
    & 2     & 214   & 1     & 2     & 1.49  & 1.78  & 1.81  & 1.47  & 1.59  & 1.00  \\
    & 5     & 214   & 4     & 8     & 1.78  & 2.46  & 2.26  & 1.83  & 1.90  & 1.00  \\
    & 10    & 217   & 6     & 15    & 2.28  & 2.86  & 2.85  & 2.20  & 2.31  & 5.29  \\
    & 15    & 221   & 7     & 21    & 1.55  & 1.91  & 1.66  & 1.54  & 1.64  & 1.00  \\
    & 20    & 222   & 11    & 30    & 1.23  & 1.38  & 1.31  & 1.19  & 1.24  & 1.89  \\
    \hline
    \multirow{5}{*}{\tabincell{c}{ consph \\ $|V|=79679 $ \\ $|E|=2963573$ \\ $d(G)=41$ \\ $cd(G)=22$}} 
    & 2     & 24    & 19    & 2     & 1.68  & 8.19  & 2.86  & 2.84  & 1.74  & 1.00  \\
    & 5     & 26    & 20    & 6     & 24.80  & 45.23  & 15.65  & 14.71  & 24.87  & 1.89  \\
    & 10    & 33    & 18    & 9     & 8.93  & 78.64  & 29.43  & 15.51  & 8.86  & 1.25  \\
    & 15    & 42    & 14    & 10    & 16.68  & 87.12  & 21.00  & 91.69  & 16.19  & 1.45  \\
    & 20    & 45    & 16    & 17    & 261.17  & 725.54  & 249.05  & 1621.17  & 233.18  & 1.84  \\
    \hline
    \multirow{5}{*}{\tabincell{c}{ C125.9 \\ $|V|=125 $ \\ $|E|=6963$ \\ $d(G)=102$ \\ $cd(G)=84$}} 
    & 2     & $\ge$ 42    & $\le$ 62    & $\le$ 46    & OOT   & OOT   & OOT   & OOT   & OOT   & OOT \\
    & 5     & $\ge$ 63    & $\le$ 44    & $\le$ 31    & OOT   & OOT   & OOT   & OOT   & OOT   & OOT \\
    & 10    & $\ge$ 91    & $\le$ 21    & $\le$ 13    & OOT   & OOT   & OOT   & OOT   & OOT   & OOT \\
    & 15    & 112   & 5     & 2     & 65.63  & 4.82  & 65.65  & OOT   & 84.24  & 6.29  \\
    & 20    & 122   & 0     & 2     & 0.00  & 0.00  & 0.00  & 0.00  & 0.00  & 1.00  \\
    \hline
    \multirow{5}{*}{\tabincell{c}{ MANN\_a27 \\ $|V|=378 $ \\ $|E|=70551$ \\ $d(G)=364$ \\ $cd(G)=350$}} 
    & 2     & 236   & 130   & 118   & 562.86  & 1299.49  & 609.94  & 609.33  & OOT   & 2.22  \\
    & 5     & 351   & 18    & 9     & 0.78  & 0.40  & 1.01  & 0.75  & 0.93  & 1.68  \\
    & 10    & 351   & 23    & 19    & 266.10  & 213.03  & 269.07  & 18.74  & OOT   & 2.25  \\
    & 15    & 378   & 1     & 2     & 0.01  & 0.01  & 0.01  & 0.01  & 0.01  & 1.00  \\
    & 20    & 378   & 6     & 12    & 0.01  & 0.01  & 0.01  & 0.01  & 0.01  & 1.00  \\
    \bottomrule
    \end{tabular}
}
\end{table}%

\subsubsection{Analysis on Graph Reduction.}
To illustrate the effectiveness of our graph reduction, especially of the second- and higher-order reduction, the running time of Maple$\mathrm{_{NGR}}$ is also reported in Table \ref{tab-demo}, column $T_{\overline{R}}$. In general, graph reduction significantly improves practical performance. When $k=2,5$, graph reduction can achieve up to 2x speedup as shown on the soc-lastfm graph with $k=5$. When $k=10,15,20$, the benefit of graph reduction becomes more evident. For example, when $k=10$, Maple solves the soc-lastfm graph in 2.25s, while Maple$\mathrm{_{NGR}}$ fails to solve it within 1800s time limit. As another example, when $k=15$, Maple solves the sc-pwtk graph in 18.59s, while Maple$\mathrm{_{NGR}}$ still runs out of time. It is also worth noting that graph reduction does not always have a positive effect. For example, Maple takes 266.10s to solve the MANN\_a27 graph, whereas Maple$\mathrm{_{NGR}}$ solves it with only 18.74s. The underlying reason is the computation overhead to perform the reduction rules.

\subsubsection{Analysis on Bounding Technique.} 
To verify the effect of our bounding technique, especially for the minimum $d$-bdd size bound, the running time of Maple$\mathrm{_{NBD}}$ is also reported in Table \ref{tab-demo}, column $T_{\overline{B}}$. For the large real-world graphs from Network-Repo and 10th-DIMACS datasets, the effect of our bound is not as positive as our reduction rules. For example, due to computational overhead, although Maple outperforms Maple$\mathrm{_{NBD}}$ on the soc-lastfm graph when $k=5$, Maple is not as time-efficient as Maple$\mathrm{_{NBD}}$ on the soc-lastfm and consph graphs when $k=20$. But for artificial dense graphs from the 2nd-DIMACS benchmark, our bound proves to be very useful. For example, when $k=2$ and 10, Maple solves the MANN\_a27 graph with 562.86s and 266.10s, respectively, whereas Maple$\mathrm{_{NBD}}$ fails to solve both of them. To sum up, our bound is more suitable for artificial dense graphs than real-world graphs. 

\subsubsection{Analysis on Branching Factor.} 
For a branching algorithm, the \textit{branching factor} is defined as the number of sub-branches at each branch.
In the worst case, the branching factor of DBDD subroutine is $k+1$.
However, this worst-case estimation is pessimistic.
In Table \ref{tab-demo}, we also show the average branching factor $\gamma$.
In general, the average branching factor changes slightly as $k$ increases. In fact, it is always much smaller than $k+1$.
We observe that when $k=2$, the branching factor on the consph and soc-livejournal graphs is $1.00$, meaning that an optimal solution can be found without branching. 
Even for C125.9, the branching factor $6.29$ is still much smaller than the worst-case value $16$, given $k=15$.
The low value of the average branching factor is achieved by incorporating the aforementioned reduction rules and the bounding technique, which indicates that our algorithms run much faster than the worst-case estimation. 

\section{Conclusion and Perspective}

In this paper, we presented new exact algorithms for the maximum $k$-plex problem and more importantly, we prvoided explanations of why these exponential-time  algorithms can be efficient in practice.
Our first algorithm is parameterized by the degeneracy gap $g_k(G)$, a parameter which is the difference between the degeneracy bound of graph $G$ and the size of the maximum $k$-plex in $G$.  We extended our result with the community-degeneracy, which gives another algorithm parameterized by the community-degeneracy gap $cg_k(G)$, an even smaller parameter.

We showed in experiments that our algorithms are competitive with state-of-the-art algorithms.
Then, we demonstrated that both the degeneracy gap and the community-degeneracy gap are very small in many real-world graphs, where our algorithms are very efficient. Finally, we conducted a correlation analysis to confirm the indicative function of gap parameters on the running time.
Altogether, this work takes an important step towards bridging the gap between algorithmic complexity theory and practical efficiency in terms of running time.

This line of research can be extended from multiple perspectives. First, there should be more theoretical or empirical investigations between the multiple parameters and the running times.  For example, one could build a systematic framework to incorporate all these parameters in one algorithm or include these graph parameters in an automatic run-time prediction tool. This has also been pointed out in the work of \cite{figiel2022correlating}.

On the other hand, empirical fast algorithms with theoretical guarantees have been pursued for years in communities like operations research and artificial intelligence. This work provides novel insights into the use of parameterized complexity for solving hard graph problems. Specifically, graph parameters like degeneracy gaps can be extended and used in other graph models, like $k$-bundle \citep{zhou2022effective,hu2023listing}, $k$-defective clique \citep{chen2021computing,gao2022exact} and so on. 

Lastly, due to the importance of the $k$-plex model, our algorithm can be used or served as a subroutine in many other applications. For example, since the algorithm also works for the enumeration of maximal $k$-plexes, we can use this algorithm to accelerate the community detection task in \citep{conte2018d2k,zhu2020community}. The code of our algorithms that we make publicly available facilitates such applications.

% Acknowledgments here
\ACKNOWLEDGMENT{%
Special thanks are given to Professor Christian Komusiewicz who introduced some important related works.
This work was partially supported by CCF-Huawei Populus Grove Fund - Theoretical Computer Science Special Project, China Postdoctoral Science Foundation under grant 2022M722815, Natural Science Foundation of Sichuan Province of China under grants 2023NSFSC1415 and 2023NSFSC0059, and National Natural Science Foundation of China under grant 61972070.}

% Leave this (end of acknowledgment)

% Appendix here
% Options are (1) APPENDIX (with or without general title) or 
%             (2) APPENDICES (if it has more than one unrelated sections)
% Outcomment the appropriate case if necessary
%
% \begin{APPENDIX}{<Title of the Appendix>}
% \end{APPENDIX}
%
%   or 
%
\begin{APPENDICES}
\section{Proofs}
\subsection{Proofs of Theorem \ref{thm-commu}}

We have the following conclusion.
\begin{lemma} 
	\label{lemma-COMMUPLEX-complex}
	Given a graph $G=(V, E)$, a fixed integer $k\ge 1$ and an integer $p$ that $p \ge 2k-1$, KPLEX$\mathrm{_{com}}$$(G,k,p)$ solves the $k$-PLEX problem in time $O(|V|^{O(1)}(k+1)^{cg})$, where $cg=cd(G)+2k-p$.
\end{lemma}
\proof{Proof.}
Let $T_{KPLEX\mathrm{_{com}}}(G,k,p)$ denote the running time of KPLEX$\mathrm{_{com}}$$(G,k,p)$. Then,

\begin{small}
	\begin{align*} 	 
		&T_{KPLEX\mathrm{_{com}}}(G,k,p) = T_{commu}(G)+ \\
		&\sum_{i=1}^{|E|}\sum_{\substack{S\subseteq N^{2+}_G(e_i) \\ |S|\le 2k-2}}  
		{\Bigl(T_{graph}(G_s)+T_{DBDD}(\overline{G_s}, k-1, t, N_G^+(e_i), \emptyset)\Bigr)},
	\end{align*}
\end{small}
where $T_{commu}(G)$ is the time of community-degeneracy ordering, $T_{graph}(G_s)$ is the time of building $G_s$ given an edge $e_i=\{u,v\}$ and a subset $S$, and $T_{DBDD}(\overline{G_s}, k-1, t, N_G^+(e_i), \emptyset)$ is the time of running DBDD$(\overline{G_s}, k-1, t, N_G^+(e_i), \emptyset)$.
Note that $G_s=(V_s,E_s)$ is the vertex-induced subgraph $G[\{u,v\}\cup S \cup N^+_{G}(e_i)]$ and $t=|V_s|-p$.

It is known that $T_{commu}(G)\le O(|V||E|)$ by \cite{buchanan2014solving}.
Define $cg=cd(G)+2k-p$. Since $|N^+_{G}(e_i)|\le cd(G)$ and $|S|+2 \le 2k$, we have $|V_s| \le cd(G)+|S|+2 \le cd(G)+2k$ and $t \le cd(G)+2k-p = cg$.
Therefore, $T_{graph}(G_s)\le (cd(G)+2k)^2$ and $T_{DBDD}(\overline{G_s}, k-1, t, N_G^+(e_i), \emptyset)\le (cd(G)+2k)^2(k+1)^{cg}$.
Therefore,
\begin{small}
	\begin{align*} 	 
		&T_{KPLEX\mathrm{_{com}}}(G,k,p) = O\biggl(|V||E|+ \\
		&\sum_{i=1}^{|E|}\sum_{\substack{S\subseteq N^{2+}_G(e_i) \\ |S|\le 2k-2}}  
		{\Bigl((cd(G)+2k)^2+(cd(G)+2k)^2(k+1)^{cg}\Bigr)}\biggr) \\
		&=O\biggl(|V||E|+\sum_{i=1}^{|E|}{|V|^{2k-2}(cd(G)+2k)^2(k+1)^{cg}}\biggr) 	\\
		&=O(|V||E|+|V|^{2k-1}(cd(G)+2k)^2(k+1)^{cg}).		
	\end{align*}
\end{small} 
which ends the proof.
\Halmos
\endproof

Finally, for each $p$ starting from $|V|$ to $\omega_k(G)$, we run KPLEX$\mathrm{_{com}}$$(G,k,p)$.  At each time KPLEX is called, we have $cg\le cd(G)+2k-\omega_k(G)$. Therefore, Theorem \ref{thm-commu} holds.

\subsection{Proof of Graph Reduction}
\proof{Proof.}
    Given a vertex set $P\subseteq V$, let $n = |P|$, $\lambda = |E(G[P])|$. Assume that $|\Delta_G(P)| < p-nk+n(n-1)-2\lambda$, but all vertices of $P$ belong to a $k$-plex $P^*$ of size at least $p$.
    
    For any vertex $u\in P$, there are at most $k-(n-|N_{G}(u)\cap P|)$ vertices that are not adjacent to $u$ in $P^*\setminus P$.
    Hence, there are at most $\sum_{i=1}^{n}k-(n-|N_{G}(u_i)\cap P|)$ vertices in $P^*\setminus P$ that are not belong to $\Delta_{G}(P)\cap P^*$.
    Then,
    \begin{equation}
    	\begin{aligned}
    		&|P^*|-n-|\Delta_{G}(P)\cap P^*| \le \sum_{i=1}^{n}k-(n-|N_{G}(u_i)\cap P|) \\
    		&= nk-n^2+\sum_{i=1}^n|N_{G}(u_i)\cap P| = nk-n^2+2\lambda
    	\end{aligned}
    \end{equation}
    In summary, we have $|\Delta_{G}(P)\cap P^*| \ge |P^*|-nk+n(n-1)-2\lambda \ge p-nk+n(n-1)-2\lambda$.
    It is clear that $|\Delta_{G}(P)\cap P^*| <= |\Delta_G(P)|$, which contradicts the assumption that $|\Delta_G(P)| < p-nk+n(n-1)-2\lambda$.
\Halmos
\endproof

\subsection{Proof of Lemma \ref{lemma_bound}}
\proof{Proof.}
Assuming that the vertices in $\pi_0$ form an independent set, it is obvious that this assumption will not increase the lower bound of the minimum $d$-bdd size. Thus we can move all vertices in $\pi_0$ from $C$ to $S$.

As $G[S]$ is $d$-degree bounded and $\pi_i$ is a subset of $v_i$'s neighbors in $C$, then at most $d - \delta_i$ vertices in $\pi_i$ can be included in $S$ at the same time.
Hence, at most $\Sigma_{i=1}^{q} min\{d - \delta_i, |\pi_i|\}$ vertices in $C \setminus \pi_0$ can be included in $S$.

To sum up, $|C| - |\pi_0| - \Sigma_{i=1}^{q} min\{d - \delta_i, |\pi_i|\}$ is a lower bound of the minimum $d$-bdd excluding $S$ in $G$.
\Halmos
\endproof
\section{Missing Charts}
\begin{figure}[htbp!]
\centering
\includegraphics[width=0.65\columnwidth]{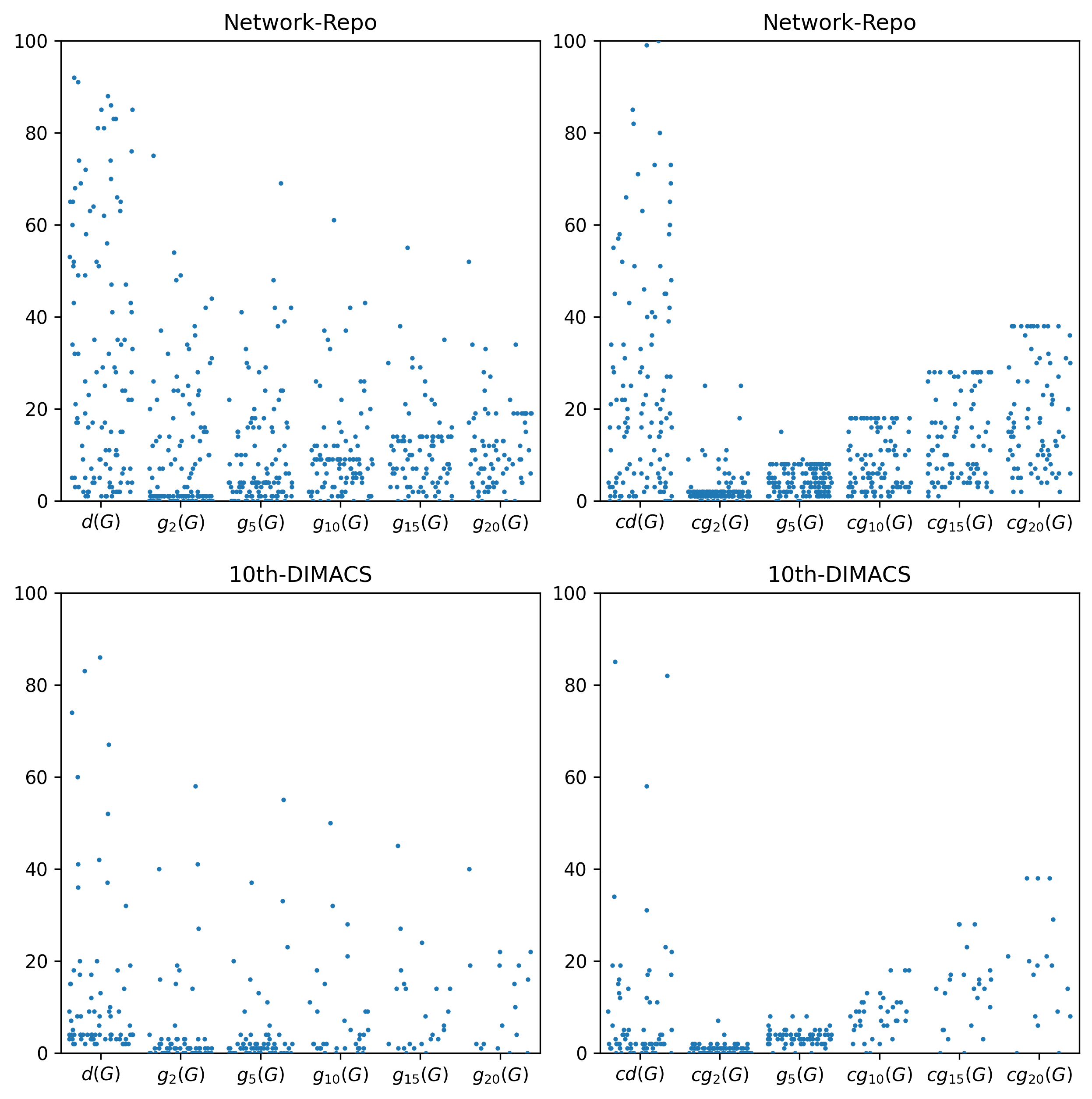}
\caption{Scatter graph of $d(G)$, $cd(G)$, $g_k(G)$ and $cg_k(G)$ for Network-Repo and 10th-DIMACS graphs, $k=2,5,10,15,20$. (Y-coordinate is cut off by 100 due to space limit.)\label{fig-scatter}}
\end{figure}
\end{APPENDICES}

% References here (outcomment the appropriate case) 

% CASE 1: BiBTeX used to constantly update the references 
%   (while the paper is being written).
\bibliographystyle{xxxxxxx} % outcomment this and next line in Case 1
\bibliography{xxxxxxx} % if more than one, comma separated

% CASE 2: BiBTeX used to generate mypaper.bbl (to be further fine tuned)
%\input{mypaper.bbl} % outcomment this line in Case 2

\end{document}